# From Reality Keys to Oraclize. A Deep Dive into the History of Bitcoin Oracles

Giulio Caldarelli
University of Verona
Department of Management
Via Cantarane 24, 37129, Verona (VR), Italy
giulio.caldarelli@univr.it

v0.2

**Abstract**

Before the advent of alternative blockchains such as Ethereum, the future of decentralization was all in the hands of Bitcoin. Together with Nakamoto itself, early developers were trying to leverage Bitcoin's potential to decentralize traditionally centralized applications. However, being Bitcoin a decentralized machine, available non-trustless oracles were considered unsuitable. Therefore, strategies had to be elaborated to solve the so-called "oracle problem" in the newborn scenario. By interviewing early developers and crawling early forums and repositories, this paper aims to retrace and reconstruct the chain of events and contributions that gave birth to oracles on Bitcoin. The evolution of early trust models and approaches to solving the oracle problem are also outlined. Analyzing technical and social barriers to building oracles on Bitcoin, the transition to Ethereum will also be discussed.

## 1. Introduction

"*That's cheating, though, isn't it?...but all of the really interesting complex contracts I can think of require data from outside the blockchain*"[1]. "Cheating" is how Gavin Andersen provocatively referred to utilizing oracles on the blockchain to run smart contracts. The idea is that, in order to be able to finally utilize the Blockchain for something above cryptocurrencies, renouncing to a degree of decentralization may be considered a fair take. Whether it is right to leave the hard-achieved decentralization and the degree to which it has to be renounced in exchange for more interoperability is yet to be defined [2]–[4].

The literature on blockchain oracles is a small niche. Two recent studies show that the total number of academic papers concerning oracles barely exceeds two hundred [5], [6]. The academic and practitioner interest in blockchain oracles rose, in fact, after the 2017 ICO hype, where hundreds of blockchain integration proposals were launched almost in every sector [7]. Since many turned out to be fraudulent or unrealistic, studies arose on the motive of their infeasibility [8]–[11]. An emergent stream of literature guided by the works of Egberts [12], Frankenreiter [13], and Damjan [14] also started to investigate the role of oracles, along with their uses, risks, and legal implications in real-world blockchains. As the general awareness of the so-called "oracle problem" increased, many other papers concerning oracle technical structure and classification emerged [2], [15]. The paper by Al-Breiki et al. [16] is one of the first to classify trustworthy blockchain oracle protocols by evaluating their security and the foundations of their trust models. Eskandari et al. [17] and Liu et al. [18] instead focus on oracles used in DeFi. The first provides a theoretical framework to classify them, while the second outlines, by gathering on-chain data, the deviation rate of the different oracle designs. Recent research by Pasdar et al. [19] instead involves a consistent number of oracles. It investigates the data type they can provide, their resistance to Sybil attacks, and their exposure to the so-called “verifiers dilemma.” The dilemma concerns the preference of the verifier to vote for the outcome that guarantees himself a reward instead of performing work for correctness.

It has to be said that since the academic literature on oracles started in 2017-18, the whole decentralized infrastructure had already shifted from Bitcoin to Ethereum and other alt-chains by that time. Therefore, the undergone studies mainly involved the infrastructure active and observable in that timeframe and onward, reflecting thus a specific philosophy and belief. However, the concept of decentralizing applications with blockchain and the use of oracles is intuitively much older [20]. Before the advent of Ethereum, Bitcoin was the leading ecosystem on which early smart contracts developers and blockchain enthusiasts experimented with decentralized applications. As data from the real world requires oracles, those had to be primarily theorized and built on Bitcoin. Although research in [12], [20] hinted that oracles on Bitcoin worked with multi-signature wallets, to the best of the author’s knowledge, no broader and further studies can be retrieved on how those protocols were theorized and built, nor how their trust model worked. Since Bitcoin is much older than Ethereum, it is reasonable to hypothesize that more than one oracle type was created and active on top of its chain.

The idea of this paper is that the oracle literature broadly misses his Bitcoin history, which should also include their origin. Due to that reason, oracles theoretical background and evolution may be biased by an investigation of projects already developed on alt-chains. Research in [5] supports the view of excessive heterogeneity and confusion in oracle definitions and boundaries. Unclarity also emerges in the characteristics of their trust models. As the oracle origin and underlying idea have yet to be defined, it is arguable that those aspects may be clarified by investigating their history further.

In the absence of dedicated academic or grey literature, the author opted for an exploratory study to investigate the oracle's origin. The data collection is therefore guided by experts in the field who were among the first to theorize and develop oracles on the Bitcoin blockchain. Their theoretical background and the history of their protocol are outlined to understand better how the oracle concept was born and how it theoretically evolved. The technical structures are also outlined and compared to show how technically the trust model evolved with every oracle. Their history and their protocols will be traced and described. Where possible, the data provided by the experts is double-checked with available written documents, repositories, and online materials, such as emails and forum posts.

The research questions of this study are the following:

1) What is the exact origin of blockchain oracles, and how were they theorized?
2) How did early developers face the oracle problem?
3) How have trust models evolved?
4) Which factors mainly contributed to the shift oracles development to the Ethereum ecosystem?

The paper proceeds as follows. Section two introduces the methodology as well as the data collection while section three outlines the findings. Section four discusses the findings and answers the research questions. Section five concludes the paper by providing hints for further research.

## 3. Findings

This section provides an overview of early oracle protocols starting from Reality Keys and ending with Oraclize. Every paragraph is divided into two sub-paragraphs in which the first retraces the history of the oracle protocol from its underlying idea to its development and difficulties faced, while the second technically describes the oracle module. Table one provides the list of experts and associated protocol/paragraph. Quotations from each paragraph (unless stated otherwise) are from the relative expert interviewed.

Table 1. Experts interviewed and dedicated paragraph.

| Expert | Protocol/Paragraph |
|---|---|
| Mike Hearn | Oracles' origin |
| Edmund Edgar | Reality Keys |
| Paul Sztorc | Truthcoin |
| Tomasz Kolinko | Orisi |
| Adam Krellenstein | Counterparty |
| Thomas Bertani | Oraclize |

### 3.1. The origin of blockchain oracles.

In 2012, during a Bitcoin conference in London, Mike Hearn, one of the first Bitcoin developers, declared that he noticed some unusual pieces of code in Bitcoin. When Mike asked Nakamoto for clarifications, he replied that they were meant to execute "contracts" at a later time. It appears that these so-called contracts were only visible to those that looked into the Bitcoin code, but proof of their existence could also be found in an Email from Nakamoto dated the 09th of March, 2011 [21]. In that document, Nakamoto defines contracts as "*by signing, an input owner says "I agree to put my money in, if everyone puts their money in and the outputs are this.*". Unfortunately, by the time Nakamoto left, the code to support contracts was insufficient to execute them fully, and other developers were not informed on how to continue their development.

To help solve the problem, Mike Hearn started a BitcoinWiki page concerning those so-called distributed contracts on the 22nd of May 2011, defining them as "a method of using Bitcoin to form agreements with people, anonymously and without the need for trust." The wiki page saw the contributions of many other developers, of which some probably used their real names and others just pseudonyms. Appendix 1 provides an overview of page contributors and contribution types.

A classic example of the contract described on the page was the promise of later payment using data from the blockchain (e.g., timestamp) to determine the exact time to unlock the money. However, apart from this, a contract such as a "will" was also described. A will contract concerns the event of death and introduce arbitrary data on the blockchain for the first time. Due to that idea, on the 25th of July 2011, the concept of Oracle was added on the wiki page by Mike Hearn, explaining that "as Bitcoin nodes cannot measure arbitrary conditions, we must rely on an 'oracle''. In the same contribution, an oracle was defined as "a server that has a keypair, and signs transactions on request when a user-provided expression evaluates to true" [22].

In the will example described in the wiki, the oracle was the third key owner of an M-of-N multi-signature wallet that signs the transaction when the condition death=true. The contract illustrated in figure 1 is meant to work as follows:

The creator of the will (e.g., grandfather for grandson) would create a transaction spending the output and setting the output to:

<oracle pubkey> <grandson pubkey> 2 CHECKMULTISIGVERIFY <hash> OP_TRUE

It means that the transaction is complete by the grandfather side, but its expendability is conditional to the script's output mentioned above. The script requires two other key owners to sign the transaction when a specific hash is verified.

The oracle then accepts the request and receives the expression and a copy of the partially complete transaction along with the output script. The oracle pubkey would be published on the oracle website, which is meant to be a trusted data source (in this case, it concerns people's deaths). Then the sentence about the grandfather's death should be in a form that the oracle can understand (e.g., a hash). Ideally may be a hashed form of the string:

has_died('john smith', born_on=1950/01/02)

The oracle then verifies if the hash of the expression matches the hash of the output scripts, and if it does, he signs the transaction. Otherwise, he returns an error.

Assuming that the grandson has already signed his part of the script, when the oracle successfully signs his part, the grandson can broadcast the contract transaction and the money claim[22], [23].

It must be noted that, in this example, the creator of the will contract decides the oracle that can unlock the transaction.

Figure 1. Multi-Sig based will Contract.

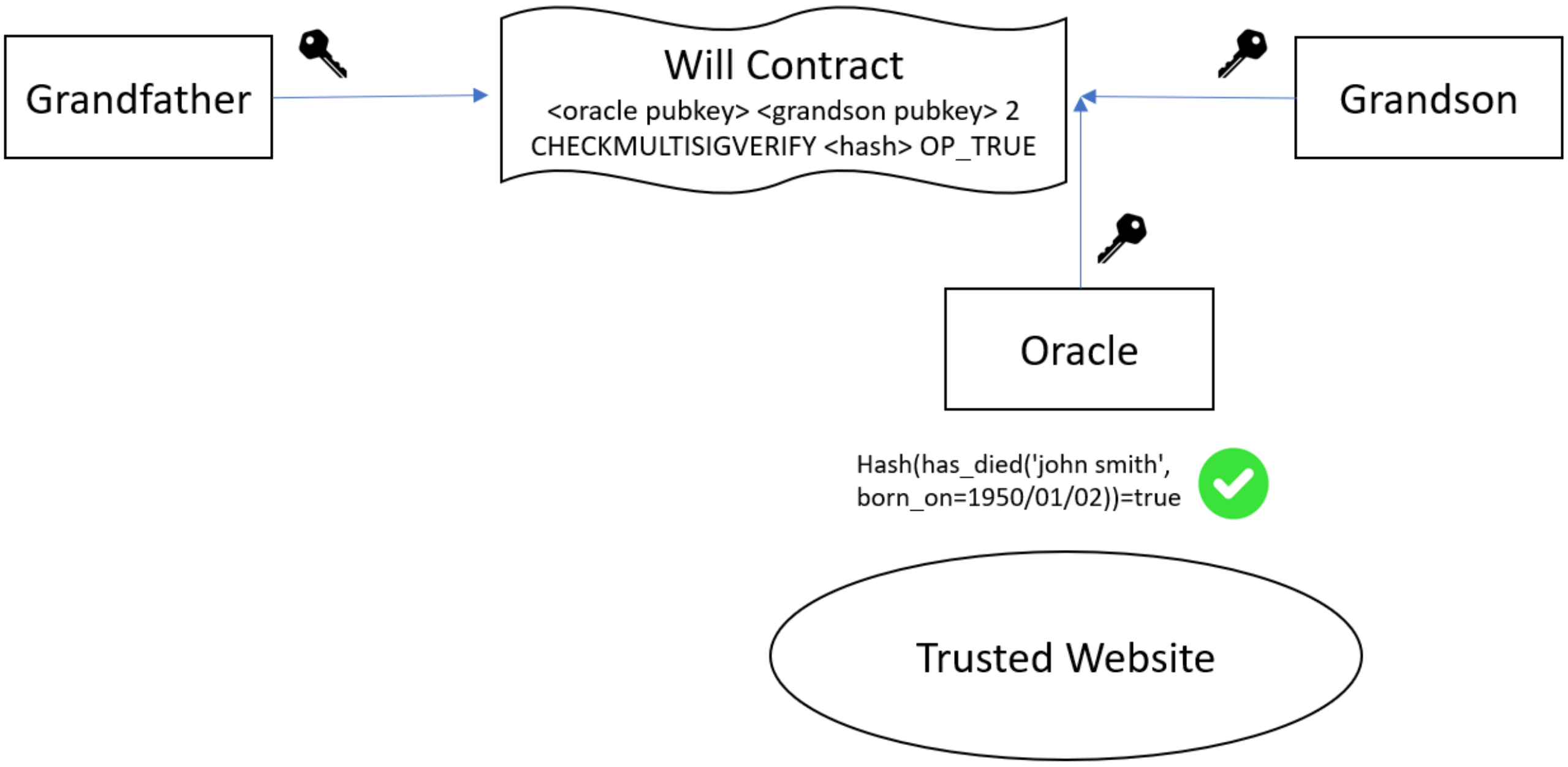

By that time, the approach was purely theoretical, and Mike Hearn had never developed an actual oracle. If the oracle was known as a black box, his efforts were toward "how to make that black box a little more transparent." As for the reason why he gave to this new concept in the blockchain space, the name oracle, he replied: "The name itself (oracle) is a bit of everything, just like contracts; all these things are metaphors. I think I used it because there is a history of using that term in the field of cryptography, and what I was developing was similar to the concept of random oracle".

Hearn further realized that a working Bitcoin wallet was necessary for these new features to be successfully implemented. In 2012-13 however, the wallet market was small and fragmented. In 2014 Bitcoin wallets were even banned by Apple [24]. Aiming for a solution, Mike Hearn started the development of Bitcoin-J, a wallet that could successfully support contracts. Unfortunately, many other difficulties prevented the effective development of contracts and oracles by that time. In Mike's opinion the following were the most problematic:

- Development Hardness – Contracts, as a new type of programming, were not easy to develop. They involved cryptography which is something programmers are not always very familiar with.
- Non-interoperable wallets - As developers were developing their own wallets, there was an evident lack of interoperability. Programs hardly worked on one wallet and rarely worked on others.
- Incomplete contributions - Many contracts were just proof of concept, not even integrated into a wallet, and often only executable from the command line. As the final integration into a wallet was difficult, people weren't just doing it.

According to Mike Hearn, apart from being a big experiment, oracles will not be used more broadly in the future. He believes that the real solution lies in the "trusted computing" area he has dedicated to after leaving the Bitcoin community in 2015.

### 3.1.1. Nakamoto "contracts" and considerations on oracles.

Considerations directly coming from Nakamoto regarding transactions involving an external state or arbitrator can be traced back to an email of the 27th of April 2009. The email discussed Mike Hearn's idea of introducing chargebacks to Bitcoin transactions [25]. In 2009 Bitcoin was at its earliest stage, and no scripts had been implemented yet. In that email, Nakamoto stated that if an agent required the possibility of chargeback, an "escrow" transaction (which was still not implemented) should have been used. Furthermore, a third party with the power to decide whether to return or release the money had to be designated. The idea was also to implement an expiration date to the escrow for the funds to be automatically returned if no options were exerted within the time limit. Interestingly, the original proposal of Nakamoto was slightly different. He proposed, in fact, an escrow system in which either the bitcoins were released or burnt. It was a sort of "kill switch" that prevented thieves to gain benefits from cheating [26]. However, the community voted against the burning mechanism, opting for a chargeback mechanism instead. The kill switch was thought to excessively penalize the buyer (and also the whole community) by permanently removing bitcoin from circulation [27].

Except for the chargeback example, the idea of a third-party arbitrator is not retrieved anywhere else in Nakamoto's writings when Bitcoin scripts were finally available. As a matter of fact, in the same email in which Nakamoto describes the idea of contracts, the provided resolution does not involve an arbitrator. In his contract based on multi-sig, all the participants are key owners with the same rights, and the resolution is attained as soon as the required subset of parties signs the transaction.

In a following communication, Nakamoto also considered the possibility of broadening the range of applications built on Bitcoin. He declared that he was planning to build a marketplace "eBay style," built into the client, but with the same mechanism of review/ratings of modern intermediary platforms. However, due to the "locked-in nature" of Bitcoin he saw it difficult for those applications to be directly built on top of the chain. Therefore, he shared the idea of utilizing other chains (e.g., sidechains) with more developer-friendly rules, but with the same miners as Bitcoins. To achieve interconnectedness, he suggested that inputs for the other chains could have been data from Bitcoin blocks (e.g., the nonce). The alternative chain to which Nakamoto refers was named "BitDNS" by that time. Still, it was only a theoretical approach, and it seems unrelated to existing projects sharing the same name [28], [29].

A few days before his last email [30], Nakamoto also commented on the possibility of Bitcoin scripts being non-stateless. His opinion was that if Bitcoin has access to outside data which may change between nodes, it could generate a fork in the chain. The exception is made for information that is always false up to a specific time and permanently true after (Timelocks). In reply to another email from Mike Hearn, Nakamoto, also provided his view concerning the involvement of trusted third parties, such as Google, in managing users' accounts. Nakamoto suggested that even in the presence of trusted third parties, contracts should be executed in a trustless way. This can be attained if the trusted third party signs the transaction before the contract is created.

As this was the last contribution from Nakamoto on Bitcoin and external data, no information could be retrieved on how he wished to put arbitrary data on sidechains.

### 3.2. Bridging Reality to the Bitcoin blockchain: Reality Keys

Fascinated by the episode of the WikiLeaks banking blockade, Edmund Edgar came across discussions about peer-to-peer currencies, eventually discovering Bitcoin in 2010. His first contribution to the ecosystem concerns the attempt to introduce Bitcoin as an official currency in OpenSim (an open alternative to Second Life). He thought that since OpenSim was a decentralized life simulator, it could have benefitted from a form of decentralized currency. While working on this integration, he came across a video of Mike Hearn from 2012 in London in which he talked publicly about the possibility of using Bitcoin Technology for real-world applications. He started following then, many discussion threads on BitcoinTalk about the need for a trusted oracle for Bitcoin in order to build real-world applications [31]. He noticed, however, that it had yet to be an official practical implementation of this idea, so he started working on his own. In principle, he was working in parallel on both the OpenSim currency and oracle for Bitcoin, but since OpenSim was not gaining attention and oracles were a more interesting subject, he decided to focus entirely on what later became Reality Keys.

The first lines of code for Reality Keys were written in late 2013, and the project was released early in 2014. Being the first official working oracle, Reality Key was not influenced or inspired by other projects. Its ecosystem was built in response to the need of that time to create a bridge from blockchain to the real world and the strict Bitcoin technical constraints. Congesting and eventually breaking the chain with these new applications was the primary concern; therefore, Edmund thought about making its oracle ecosystem ultimately work off-chain. Furthermore, given the technical limitations of Bitcoin and to adhere to the available scripts, the oracle was set to answer only binary (yes/no) questions.

At the first implementation of Reality Keys, the research for the data and the publication of the correct answer were done directly by the project team, eventually by Edmund himself. However, the users who made the questions knew the data source from which the information was taken. Therefore, despite the system being relatively centralized and not automated, there was a certain degree of transparency. In this regard, however, Edmund specified that although his specific system design was centralized, he hoped the whole blockchain oracle ecosystem would eventually be decentralized. He expected, in fact, many other competitors to show up in the short term. Therefore, if Reality Keys was just one of the available oracles, users could freely select among the most trusted and reliable alternatives.

When the oracle system was ready and running, it served different kinds of requests, from bitcoin prices to soccer scores. There was, however, not a specific application managed by Reality Keys until they built a sponsorship integration. This used the RunKeeper API to promote walks and marathon-related events. However, personal challenges concerning walks and runs with humanitarian aims were also sponsored. Someone could, for example, challenge himself that if he doesn't run a certain amount of kilometers by a specific date, he has to send some BTC to a charity. Thanks to a system of APIs, Reality Keys could provide information on whether the user has reached his goal.

With these new implementations, the team also had to face new challenges. The integration with RunKeeper required, in fact, a dedicated website for the application. The user was then supposed to generate a key, and the website should have been able to perform a transaction with that key eventually. Since a working wallet such as Metamask was not available for Bitcoin, as well as tools for coding, the whole implementation should have been written from scratch. In the end, they managed to complete the website, but as Edgar declared, "this is very hard, and this is very hard to do securely." Unfortunately, the absence of a wallet supporting contracts and developing tools remained almost unchanged on Bitcoin. Since the system was inflexible, there was not much demand for contracts, and since there was not much demand, the interest in building those applications was eventually scarce.

The system then remained almost unchanged till the advent of Ethereum, with only binary questions available, but improvement attempts were made on the range of available data types. He added FreeBase as a data source for Reality Keys. Freebase allowed for a very wide variety of queries to be made using the structured data system run by Google.

When Ethereum started, Edmund went to Devcon 1 (Nov 2015), where he described how Reality Keys could be used in Ethereum. Being entirely built off-chain, Reality Keys was not wholly tied to Bitcoin and could be implemented on Ethereum immediately. At Devcon 1, Edgar also met Thomas Bertani from Oraclize.

From that point onward, the development of Reality Keys switched to Ethereum, mainly for three reasons. First, the Freebase website was shut down by Google. Since it constituted one of the primary Reality Keys sources of data, it eventually affected its overall utility. Second, the block size war's outcome made Bitcoin more expensive to use for contracts, ultimately decreasing demand for Reality Keys service. Lastly, the implementation of Ethereum could have allowed the switch from a yes/no based platform to a system of signed data of any type to be directly used on-chain.

Since the platform radically changed, the project was rebranded first to Realitio and finally to Reality.eth. Besides the technical differences, the new Ethereum version also had a different theoretical approach. What Reality Keys offered on Bitcoin was simply a bridge between real-world data and on-chain contracts. Therefore, it grabbed existing data from trusted sources (e.g., Freebase) and made it available.

However, the team realized there was a need for data that didn't exist anywhere. Therefore, what Reality.eth was meant to do on Ethereum was to provide data that could not be pulled from APIs or websites. The philosophy of the platform switched then from delivering data to creating data. Two factors mainly drove the design change:

1) A trusted data source (API) could not be found for some specific applications.

2) Other projects become specialized in the bridging process (e.g., Oraclize), and it was not helpful to provide a similar service.

The project then evolved to its current version, in which it can answer any human language question.

### 3.2.1. How Reality Keys (Bitcoin) worked.

The first thing to consider to understand Reality Keys' mechanics is that since it was an off-chain oracle, there was no direct interaction with Bitcoin. However, the way it works was inherently influenced by Bitcoin's technical constraints and the available scripts by the time it was designed. The following example provides an overview of how Reality Keys could be used as an oracle on Bitcoin.

Consider having two agents, Alice and Bob, who wish to bet on Bitcoin price. Alice bets that by the 01st of June 2014, the price of BTC would reach or exceed 400$, while Bob, on the contrary, bets that by the same date, the price of Bitcoin will be lower than 400$.

The agents can solve the bet themselves or entrust it to a third-party oracle such as Reality Keys. If they decide to use Reality Keys oracle, they must make a simple binary (yes/no) question on the oracle website asking whether, by the 01st of June 2014, the bitcoin price is above/equal to 400$, which corresponds to Alice's bet. If the oracle replies no, then obviously, Bob wins the bet.

Both agents know how and from which source, Reality Keys, draws the answers and trust both the source and the Reality Keys project. Otherwise, they would freely opt for another contract resolution method.

Reality Keys creates two key pairs (public and private) for both yes and no answers. The two public keys are then published on their website. When the selected date comes (the 01st of June 2014), the Reality Keys system checks the price of Bitcoin on the proposed data source, but the result is published in two stages.

First, the system automatically publishes the results (not the key) on their website and waits for an objection period. During this objection period, an agent can ask for a “human check” of the results, offering a tip of 10 mBTC [32]. Once the objection period has elapsed, the team then publishes the correct private key and deletes the private key corresponding to the false outcome.

From a technical point of view, the role of Reality Keys ends with the publication of the correct private key. However, it is vital to understand what happens or can happen between the publication of the public key and the private key on the Bitcoin network.

Although some examples were offered on the Reality Keys website, there was actually not a specific or “standard” way of implementing their oracle service. The choice of implementing a standard or non-standard multi-signature transaction or using a script (P2SH) along with any specific conditions was totally in the hands and responsibility of the users. Depending on the selected choices, different costs, technical difficulties, or security standards would have been obtained for which Reality Keys was not responsible.

One of the few still available demo scripts (realitykeysdemo.py) implements Reality Keys creating a conditional contract on the outcome of the oracle using pybitcointools [33]. The mentioned commands refer explicitly to the script described in the repository. From the user’s side, the steps are as follows:

1) Alice creates a key pair with the command below and sends the public key to Bob. She then funds her address using any Bitcoin client. Bob does the same.

./realitykeysdemo.py makekeys

2) Alice and Bob, register a Reality key and get the ID <reality_key_id> from the URL.
3) In case one of the two parties (Alice or Bob) disappeared before completing the transaction, the other party could get the money back from the temporary address with the command

./realitykeysdemo.py pay <address> -a <amount> -f [<fee>]"

4) Alice creates a P2SH address spendable by combining (Alice key + reality key-yes) or (Bob key + reality key-no). Afterward, she creates a transaction spending the contents of both her and bob temporary address to the P2SH address, using her private key. The following output is then sent to Bob for him to sign and broadcast.

./realitykeysdemo.py setup <reality_key_id> <yes_winner_public_key> <yes_stake_amount> <no_winner_public_key> <no_stake_amount>"

5) When Bob receives the partially signed transaction, he recreates it to check if the output is the same. If everything is as expected, he signs the transaction and broadcasts it.
6) When the result is issued, whoever wins the bet, Alice or Bob, can execute the following script to unlock the funds from the contract and send them to another address of their choice.

./realitykeysdemo.py claim <reality_key_id> <yes_winner_public_key> <no_winner_public_key> -f [<fee>] -d [<destination_address>]

The procedure described above is now deprecated and is no longer available due to the transition to Reality.eth. The new oracle works under different logic and premises, and being not tied to Bitcoin anymore, its analysis goes beyond the scope of this research.

### 3.3. Enabling prediction markets on Bitcoin with Truthcoin

On 26th November 2012, the Commodity Futures Trading Commission (CFTC) claimed that Intrade.com, a prediction market platform, was interfering with CFTC’s role to police market activity and protect market integrity [34]. Probably in response to this, on the 23rd of December (same year), Intrade.com closed all U.S.-based customer accounts. On 10 March 2013, InTrade ceased all operations worldwide [35].

To Paul Sztorc, an expert in prediction markets, it was disappointing but not unexpected. Paul was interested in Bitcoin at the time and resolved to find a way to leverage Bitcoin technology to launch an open and uncensorable prediction market. Other than the markets themselves, it would have included a new Peer-to-peer "oracle" to resolve them without trusted third parties.

At that time, RealityKeys was a reliable oracle in development, but Sztorc was concerned that its system could have been manipulated for information of high value. Alternatives based on multi-sig were also not viable in the long run: "I was convinced that multi-sig was not the solution to the oracle problem. If the oracle problem is like sending a man to the moon, using a multi-sig is like trying to do it with a catapult". Emerging projects also were more oriented to the idea of data feeds, therefore, to a constant update of data to the blockchain. Sztorc's approach was, however opposite: "we don't need a data feed, we don't need a frequent check…we only need to check if some information is true at a certain point".

For that reason, although in principle not interested in oracles, he developed his own (Truthcoin) by the end of 2013, publishing the first version of the whitepaper in early 2014. The Truthcoin whitepaper and the project itself were influenced by what was called the "blocksize war", a fierce dispute between Bitcoin developers on the block size growth and the emergence of numerous alternatives to Bitcoin of dubious value. As the intention was to formalize the Truthcoin idea and then to have some other group do the actual development, Sztorc never launched the project. He could not take the risk of his ideas being manipulated by charlatans or his project being erroneously labeled as a scam.

Therefore the Truthcoin whitepaper was written in a highly scrupulous detailed way so that debates such as the block size war could never happen on his project. Furthermore, to alienate any association with alt/scam coins, he strictly adhered to Bitcoin and Nakamoto's ideas. In light of this, Truthcoin was planned to be developed as a Bitcoin sidechain.

Besides being proposed by Nakamoto, sidechains were also valuable because they allowed the avoidance of using complex Bitcoin scripts. In the script design of prediction markets, "each bitcoin transaction would be like an enormous computer program". The script's length is due not only to the application's data but also to the information on how to digest that data. Although better programmable, the same limitations would have been encountered using an all-purpose blockchain like Ethereum. A dedicated sidechain was then seen as a better solution since it already knows how to process the data, and only needs minimal inputs to code each user action. All the required code is preloaded, and the full node already knows where to find and how to process the incoming data.

Although theorized and discussed in 2010, with Nakamoto's help, sidechains were still underdeveloped [29]. The first practical idea of a two-way peg sidechain was discussed in December 2013 by Luke Dashjr. Along with others, they released the Blockstream whitepaper in October 2014, a system to enable blockchain innovation via pegged sidechains [36].

Sztorc started to develop a sidechain concept ("Drive Chain"), in 2014, but being alone and aware of the work of Luke Dashjr he decided to focus on other aspects of Truthcoin project, hoping to use Blockstream sidechain once completed. Therefore, inspired by the work of Robin Hanson, he refined Truthcoin logarithmic market scoring rules (LMSR) [37]. Concerning Hanson's contribution, Sztorc stated: "each buy and sell happens unilaterally and atomically, so it was perfect for the blockchain." Apparently, LMSR also inspired what is now called Automated Market Makers (AMM) on Ethereum [38].

In 2015 the Truthcoin software was almost complete, but unfortunately, by that time, it was clear that the Blockstream sidechain project was not going to succeed in the short run. Therefore, Sztorc switched again to the development of a bitcoin sidechain, "Drivechain", (spelled without a space this time) of which an advanced version was published in November 2015, so that it could have been beneficial not only for his now rebranded project Hivemind but also for the whole Bitcoin community.

### 3.3.1. How the Truthcoin Oracle works.

In its whitepaper, Truthcoin is described as a "proof-of-work sidechain that collects information on the creation and state of Prediction Markets (PMs)" [39]. The Truthcoin protocol exploits the concept of "salience". Oxford dictionary defines salience as the quality of being particularly noticeable or important [40]. Salient information is something that should be well-known by anyone. The solution proposed in Truthcoin to achieve salience is based on time. The idea is that information is certain and true after a certain amount of time. Instead of providing a piece of information as soon as it is known, the idea is to provide it at a point where it is undoubtedly certain. Technically it is organized as follows:

Two coins are present on Truthcoin, CashCoins (CSH) and Votecoins (VTC). CSH is pegged 1:1 to bitcoin and allows users to create, buy and sell PMs shares. Votecoins represent each user's reputation, are tradable, and pay dividends over time. VTC allows users to vote on PMs decisions and collect PMs fees. The ownership of VTC can only change by the effect of voting activity. In the whitepaper, the totality of voters is referred to as a "corporation". The concept behind this is that the reputation of the entire system is more relevant than the single reputation of each individual. Votecoins are not mined but are proportionally shared among voters in a way that if someone acquires some Votecoins, someone else has less, as its total amount remains constant.

Two types of decisions are supported on Truthcoin, Binary (0,1) and Scalar (Xmin, Xmax). A third state (.5) identifies decisions that are non-resolvable or confusing. Four entities are present on the platform which are:

**Authors**: users that create a prediction market and provide initial liquidity. The difficult work of the author lies in finding a market that may attract many users and identifying a decision that will be well-known to the voters after a specific time.

**VoteCoin Owner**: user that votes on a decision. Their main task is to maintain or increase their reputation.

**Traders**: users that trade on any PMs, are the customers of the platform.

**Miners**: Those who mine blocks on the sidechain. Being a Bitcoin sidechain that allows merged mining (Hash reuse), Bitcoin miners could mine sidechain blocks at a negligible cost.

The resolution of a market and a decision on its "true" outcome described in picture 2 is as follows.

An Author adds a decision, specifying the topic and the time of resolution. Then waits for the transaction to be included in a block. When a decision is added, the author can also add a market providing initial liquidity. Then he waits for the transaction to be included in a block. When a market is added, trading begins. The market can be advertised so that users buy and sell the shares of the different market states (such as "yes" and "no"). Eventually, the event occurs and becomes "observable". After this, when the time specified by the creator has passed, the decision is considered "mature". The set of all mature decisions is called a "ballot". Staking their tokens, owners of Votecoins are called to vote for all the decisions in the ballot, and when votes are revealed, Votecoins staked are frozen. The decision is then resolved according to the consensus algorithm, which also reallocates VTC. After a decision is resolved, a waiting time of a week starts. Once the waiting time expires, another phase starts in which Miners can veto the "resolved Ballots". If more than 50% of the blocks of this period veto the ballot, then all the decisions inside the ballot must be re-voted. When all the above-mentioned phases are concluded, the redemption phase starts in which all the winning shares are given a price, and users can redeem them for CSH.

Crucial for the oracle's good outcome is the creator's role, which must choose an event whose outcome should be "salient" at a certain point in time and voters that must correctly predict the answer of the majority of voters. If an event is not salient, voters will not be able to vote, leading to an unresolved market. Otherwise, if voters are incapable of coordinating, they will be slashed off their tokens. Failing to report or report outside the accepted value range results de facto in a slash. The amount of token slashed is proportional to the distance of the reported value to the one that is identified as the true outcome.

Finally, the Truthcoin ecosystem can be broken if someone obtains 51% of the corporation. That means being able to control 51% of the system's economic value, which is considered unlikely to happen.

Figure 2. Truthcoin market resolution phases [39].

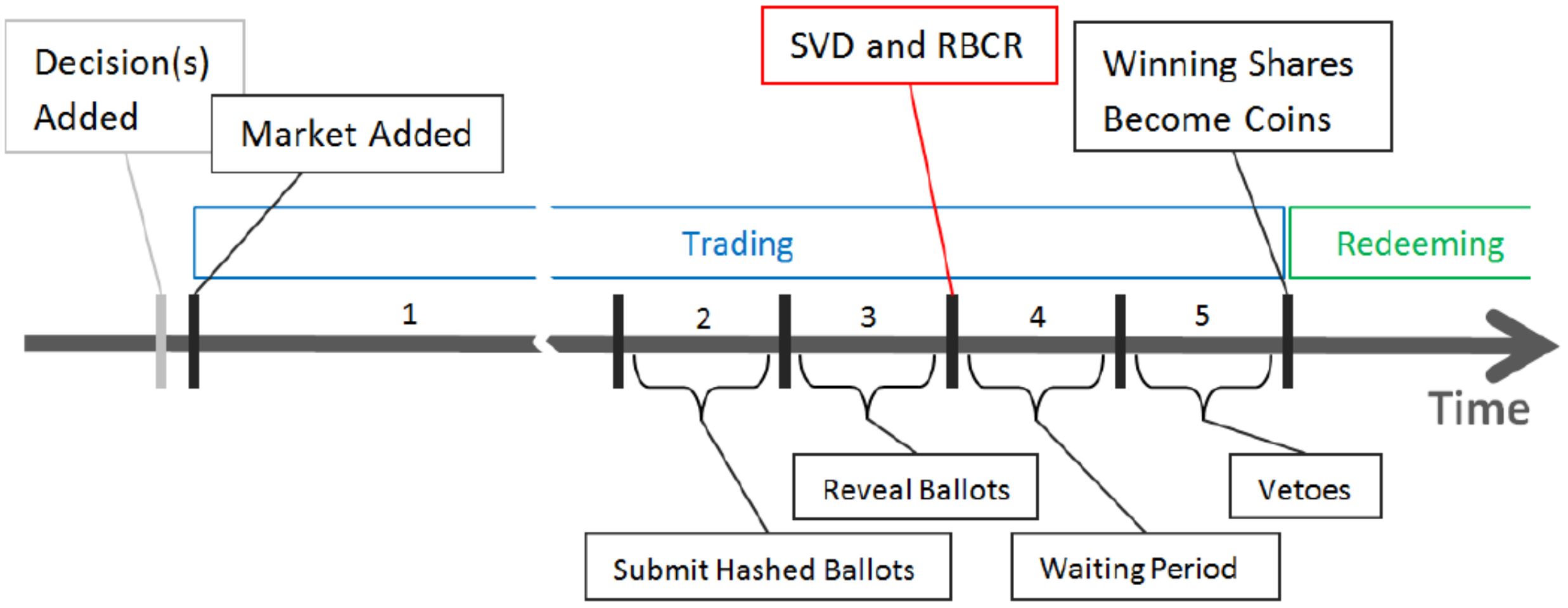

### 3.4. Multiple independent oracles on Bitcoin: Orisi

Tomasz Kolinko has been interested in Bitcoin since its early days in 2012. He had the idea of launching a stablecoin that could be transacted on the Bitcoin network. However, to build a stablecoin, a data feed was necessary to constantly update its exchange rate. By that time, the only available and known oracles were Reality Keys and Truthcoin. Reality keys was an already operating and reliable oracle project. However, in its early versions, it did not offer the possibility of a data feed, and it was mainly oriented to a one-off event such as election results. The other oracle, Truthcoin, was still under development, and similarly to Reality Keys, it was more oriented to one-off events rather than data feeds.

For that reason, he decided to develop his own decentralized oracle Orisi, based on multi-signature. He also added an entry for this concept, in the “contract” Bitcoinwiki page on the 09th of June 2014. In his idea, the oracles must have been trusted entities from the financial world (e.g., banks and other financial institutions) sending data about real-world asset prices. So, instead of adding only an oracle key, he wanted to add multiple keys to make the oracle as decentralized as possible. However, being decentralization based on multi-sig, technical limits prevented, the number of oracles to be large as Kolinko had initially in mind.

M-of-N multi-sig protocol is in fact not entirely customizable, and there is a limited key combination from which users can choose. The standard multi-signature was in fact two out of three keys, and more complex ones allowed a maximum of up to 15 keys. This structural limitation made it impossible to add more reporters to the oracle, therefore dramatically limiting the functionalities of Orisi.

Unfortunately, also another issue was encountered by Kolinko in the development of Orisi. The scripts of the Orisi oracle were in fact hardly to be mined, for two inherent reasons.

1) First, there was an economic disincentive to mine Orisi transactions because the scripts were larger than usual. As they occupied the weight of many simple Bitcoin transactions, miners could have collected more fees by selecting other transactions instead of Orisi’s. Therefore, although a bit more expensive it was unlikely for miners to mine the script voluntarily.

2) Second and most important, by that time, scripts were not a common transaction type. Pay-to-script-hash was introduced in 2012, and part of the community was not keen on inserting scripts on Bitcoin. Many wanted to keep it as a payment system only. Miners feared that processing scripts on the Bitcoin network could have broken the chain and altered the payment system. Therefore, the scripts were just not selected by the miners and left in the mempool. If, after some time, the transaction was not mined, it was automatically rejected.

It is not generally well known, but transactions on the Bitcoin blockchain are not only divided between valid and invalid. Indeed, a valid transaction is a transaction that complies with the rules of the protocol and whose output is less or equal to its input. A transaction is instead invalid if it violates protocol rules or has an output higher than its input. However, among valid transactions, miners can exert a sort of "veto" for which they can arbitrarily decide which transaction to put in their block. Censorship resistance is however guaranteed since for a miner that refuses to put a transaction in a block, there will be others that will insert the transaction for that to be eventually mined. The chance for all miners to collude to reject a specific transaction is ideally remote. However, there are also transactions that were considered as "non-standard" (e.g., multi-sig above three keys), and, although perfectly valid, were unlikely to be mined [41]. Even though the miners did not deliberately collude to reject those transactions, they were so unusual that they naturally decided not to include them. It was a sort of Schelling point [42].

Given the fact that Orisi scripts were something quite new on Bitcoin, unfortunately, despite being totally legitimate, Orisi transactions were generally not mined. To be able to have them mined, the Orisi team had to search for an agreement with a mining pool. Explaining the potential of the oracle and the underlying stablecoin projects, they managed to involve Eligius Pool which had around 7% of the Bitcoin hashrate (as of June 2014). Having 7% of the hashrate meant that Orisi transaction had 7% chance to be mined, which resulted in one transaction every 8-15 blocks on average.

Finally, due to the multi-signature limitations and the difficulties in including Orisi transactions into blocks, the Orisi project was abandoned. The multi-signature limits prevented more honest and trusted oracles to join the project, and the frequency of updates (every one or two hours) made it impossible for Orisi to serve as a price feed for a stablecoin.

### 3.4.1. How Orisi oracle worked.

Although the initial idea was to create a price feed, the Orisi whitepaper showed a far more ambitious project. The website shows active or planned support for, timelock verdicts, BTC price feed, website Boolean/integer, dedicated feed (e.g., weather feed), and arbitration support. The key innovation of Orisi oracle compared to previous proof-of-concept and proposals based on multi-sig, was to add a "set" of independent oracles. The idea was that it was difficult to bribe more than half of the oracles, and being different entities, they would have implemented different hardware and applications, thus reducing the chance of being all hacked. A list of trusted oracles was proposed on the platform website but other trusted nodes could also be selected by the users. Also, the protocol implemented a "bitmessage" protocol instead of direct IP communication to protect the identity of oracle nodes and to prevent spam, thanks to a proof-of-work mechanism [43]. The majority of oracles needed to agree to perform a transaction, and given the multi-signature limitations, they could be a maximum of 8 out of 15, in theory. However, in practice, they could not be all oracles. For example, If a multi-sig wallet is made of 4 of 7 oracles, then all the oracles may decide to send the funds to an arbitrary address. Therefore, to have seven oracles, at least an 8-of-11 multi-sign address should have been created, of which four keys should belong to the agents. Thus 1+(m of n) is turned into (n+1 of 2n-m+1). That is another factor that, given the limit of 15 keys, further constrained the usability of Orisi.

The following example (Figure 3), better clarifies how Orisi differentiates from previous oracles based on multi-sig.

Alice promises Bob that if candidate A wins the election, she will give him 10BTC. Both agree that the condition for the payment would be that on a specific website is declared the election of candidate A. Then they both agree on a set of 7 oracle nodes. Alice should then deposits 10 BTC in a multi-sig wallet that is considered a “safe” until oracles decide to forward the funds to Bob or to return them to Alice (in case candidate A loses the election). In order to do so, Alice creates an “unlock” transaction to forward the funds from the safe and pays the fees to the oracles and the Orisi project. Oracles then verify the transaction and the validity of the request. If valid they add the transaction and notify the agents. If all the oracles acknowledged the validity of the transaction, Alice transfers the 10BTC to the wallet for the contract to be finally active. If candidate A wins the election, then as soon as oracle nodes notice the information on the website, they sign the transaction that is also broadcasted through Bitmessage. Once enough oracles sign the transaction Bob can also sign the transaction with his keys and broadcast the transaction to the Bitcoin network to finally unlock the funds.

Figure 3. Multisignature contract with multiple independent oracles [44].

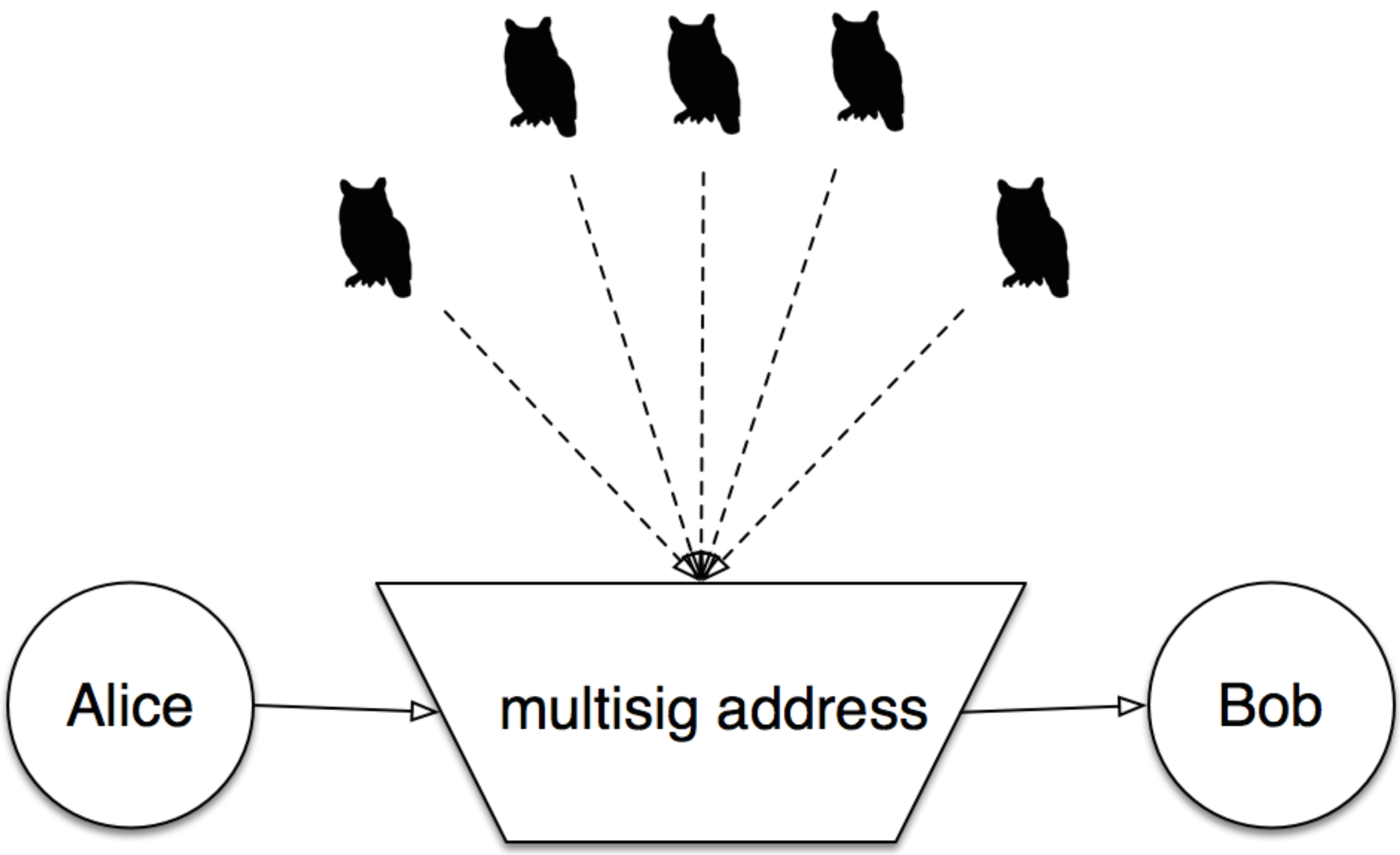

### 3.5. Bitcoin oracles through meta-chains: Counterparty

In 2012, J.R. Willet published “The second Bitcoin Whitepaper.” The document theorized the launch of Mastercoin, whose idea was to use Bitcoin as a protocol layer, on top of which higher-level protocols could be built [45]. Instead of having multiple blockchains, the Bitcoin blockchain could have been used as a foundation layer to launch new currencies also with experimental new rules [46]. Willet is also well known because to fund Mastercoin, he launched the first-ever ICO in 2013, raising 4740 BTC (nearly $500K at that time) [47]. Mastercoin (rebranded to Omni in 2015) obtained discrete attention, but the protocol development was slow at its launch despite the hype and the successful ICO. When Mastercoin was launched, the only alternative to extend Bitcoin functionalities was colored coin. Colored coin protocol was limited and

had no support for oracles; its sole purpose was to tokenize assets on a blockchain such as Bitcoin. Therefore, understanding the potential of Mastercoin, but disappointed with its slow development, Adam Krellenstein, along with Evan Wagner and Robby Dermody decided to develop a new protocol on Bitcoin called Counterparty inspired by Mastercoin premises. In particular, the goal was to be able to use information that originates off-blockchain, to produce betting, gaming, or other financial instruments on Bitcoin.

Unlike other projects, Counterparty was launched already with full functionalities on day one. Features were added quickly except for one, as the chief developer said: “the most difficult thing to add was the decentralized and trustless gaming in the form of rock paper scissors. But we had a decentralized exchange on day one, and it was already working when we launched it”. In line with other projects at that time, Counterparty was announced on Bitcointalk. Still, the developer team decided to stay anonymous at the beginning and opt for a Proof-of-burn to launch their currency (XCP), de-facto renouncing to raise any money for their project. The reasons for those choices were mainly the following:

1) The choice reflected what Nakamoto did with Bitcoin, staying anonymous and renouncing to any reward for his project.
2) There were high concerns for the legal implications of raising capital with cryptocurrencies, “which turned out to be not very serious.”
3) There were personal concerns about the project's development and how it could have turned out, as unforeseen events may have damaged personal reputation.
4) Replicating bitcoin issuance (electricity consumption), they wanted to burn resources (bitcoins) instead of transferring resources.
5) They were against raising capital for a project in the alpha-beta stage. “We didn’t want to raise money during the development as we thought it was dishonest.”

Initially, the proof-of-burn was supposed to work by consuming bitcoin as fees for the miners, de facto not really destroying bitcoins. However, many community members on Bitcointalk argued that miners could have exploited their position to produce Counterparty tokens unlimitedly. Understood that it was an actual threat, Adam Krellenstein decided to change the burning mechanism and made it by transferring BTC to an address whose private keys were unknown (e.g., an impossible vanity address), de facto making them permanently unspendable.

Above the hardships of developing a project without funding, Counterparty suffered the effect of a dispute labeled the OP_Return war [48]. In order to add the required transaction data, Counterparty needed an OP_Return size greater than the 40 bytes the Bitcoin Core developers set in the official v0.9.0 release [49]. Counterparty utilized the shrunk OP_Return feature, but the limited size also forced them to use others, such as multi-sig, to make their protocol work. Multi-sig was designed for features such as escrow payments, but the second signature could be leveraged to store data instead [50]. However, this workaround contributed to drawing the attention of the opposing faction of the OP_Return war.

From March 2014, in fact, Luke Dashjr, a Bitcoin core developer and owner of a mining pool, started to filter (eventually without success) all Counterparty transactions. As Luke declared, this censorship's motivation was to prevent exploitation of network resources by Counterparty [48], [51]. However, although probably beneficial for Bitcoin nodes, this decision was criticized because Luke was also a co-founder of Blockstream, a major Counterparty competitor [52], [53].

Although OP_Return size was increased at a later date, the main history and development of Counterparty, however, was inevitably affected by this limit [48]. Furthermore, above the constraints resulting from the reduced payload size, the most significant consequence of this debate was the fear of censorship of the Counterparty protocol. The widespread climate of uncertainty prevented developers from building on Counterparty, further impacting its development and competitiveness. It is to be noticed that also Ethereum, the second biggest network after Bitcoin, was negatively affected by OP_Return limit in its development, as Vitalik Buterin argued on social media. Although some considered it an overclaim, Vitalik declared that the original idea of Ethereum was a "counterparty-style metacoin on top of primecoin. Not Bitcoin because the OP_RETURN wars were happening"[54], [55].

Crucial in the history of Counterparty was also the advent of Ethereum. The main innovation of Ethereum was not smart contracts, de facto already available with Counterparty, but the virtual machine along with the language to write smart contracts. Counterparty had smart contracts, but only those written and supported by its developers. It was possible to code more smart contracts, but every application built had to be part of the protocol. Ethereum on the other hand had an extensible infrastructure that allowed anyone to write their own smart contracts. Only the language was part of the protocol and applications could be deployed on top of it: "*It is a more elegant and flexible system…but ultimately does the same thing*".

Aware of the value of EVM, it was ported to Counterparty (EVMParty) so that Ethereum smart contracts could be run on Bitcoin via Counterparty [56]. However, an official version was never released due to multiple factors. At the beginning, there was the idea that the user base would have been minimal, since few people were building on Ethereum and most of the applications were still on Bitcoin. After, when developers started to move to Ethereum, it was clear that Bitcoin could not compete in the smart contract field. First , contracts would have been slow due to Bitcoin block time even if more user-friendly with the introduction of EVM. Secondly, due to the block size war, the price of Bitcoin increased while Ethereum was very cheap. Therefore, nobody would have preferred Bitcoin to build contracts.

As Nakamoto did with Bitcoin, Adam Krellenstein left the Counterparty project to the community in late 2014. To date Counterparty, it is still an active project on Bitcoin and wields the same structure and premises as when it was built.

### 3.5.1. Counterparty Oracle Module explained

Counterparty is a meta-chain that runs on top of the Bitcoin blockchain. A meta-chain is a chain in which transaction data is contained on another chain called master-chain. Meta-chain transaction data runs only after Master-chain transactions are complete [57].

Counterparty transactions are Bitcoin transactions but with extra metadata in them. If a blockchain is a book, and blocks are pages, a Counterparty software writes information in the margin of those pages. Intuitively Bitcoin software ignores that extra data; therefore, specific software is needed to read it.

The Counterparty protocol's idea is that when someone signs a transaction with Bitcoin, he adds some metadata to the transaction. Then the content of this metadata is verified by all Counterparty users to ensure the transaction is valid. The architectural pattern is called state machine replication.

Let's assume that Alice wishes to transfer 5 XCP to Bob. She will then sign a Bitcoin transaction in which OP_Return (or OP_Multisig), declares the willingness to transfer 5 XCP tokens to Bob. Since Counterparty is a meta-chain, the related Bitcoin transaction will always be confirmed as long as Alice pays the necessary transaction fees. Therefore, if Alice decides to spend 5 XCP and only owns 2, the transaction on the Bitcoin blockchain will be confirmed anyway. The corresponding counterparty transaction will instead be marked as invalid.

The Counterparty transaction data is retrievable with a block explorer, but it is encoded and appears in a format such as the following:

�������6U��A��q��V��:}!mq�����m_j�-k?ee

The format is not human-readable and must be decoded by the Counterparty engine to be read and digested. The string, in fact, need to be deobfuscated with the ARC4 Cypher and verified if it starts with CNTRPRTY (first 8 bytes). From the 9th byte, information on the transaction type (send, broadcast, issuance) should be retrieved, followed by the specific transaction data. The following scheme outlines the deciphered content of a Counterparty transaction data chunk [58].

```
434e545250525459|FFFFFFFF|xxxxxx...
       |            |        |
       |            |        └──── this data is different for each transaction type.
       |            |
       |            └──────────── the transaction type identifier (4 bytes)
       |
       └──────────────────────── the string CNTRPRTY (8 bytes)
```

Once deciphered, the transaction will be digested by the Counterparty engine, which also verifies its validity.

Being able to inject extrinsic data into the blockchain, the Counterparty engine may already be considered an oracle for Bitcoin. As explained before, however, on top of Counterparty, applications such as prediction markets or decentralized exchanges can be built that further require data from the outside. Data such as a price feed for a decentralized exchange is injected into the protocol thanks to the "broadcast" transaction type. A broadcast message publishes textual and numerical information, along with a timestamp. A series of broadcasts from the same address is called a "feed." Intuitively, the timestamps of a feed should increase monotonically [59]. On the Counterparty explorer (xchain.io), users can leave feedback for the address that publishes feeds, with also some comments. Figure 4 provides an overview of what a broadcast price feed shows (3a), the details of the transaction (3b), and feedback (3c).

Figure 4a: Broadcast information (BTC-USD price feed)

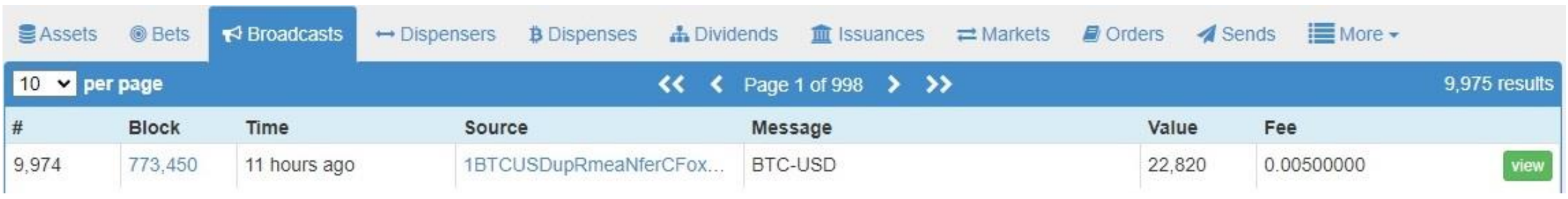

Figure 4b: Broadcast transaction details

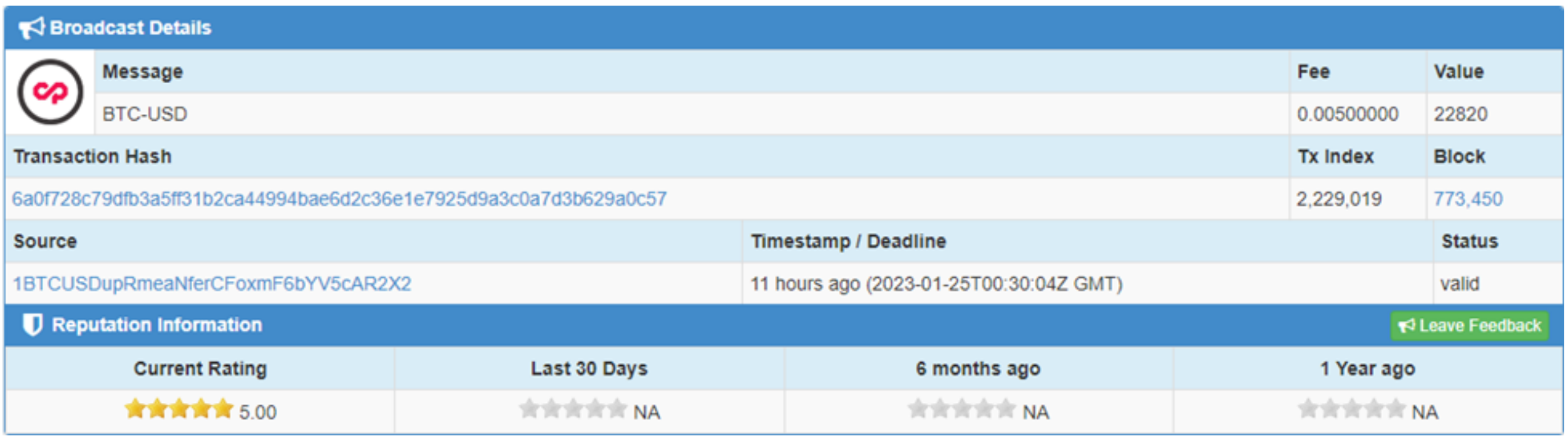

Figure 4c: Rating of the broadcaster address

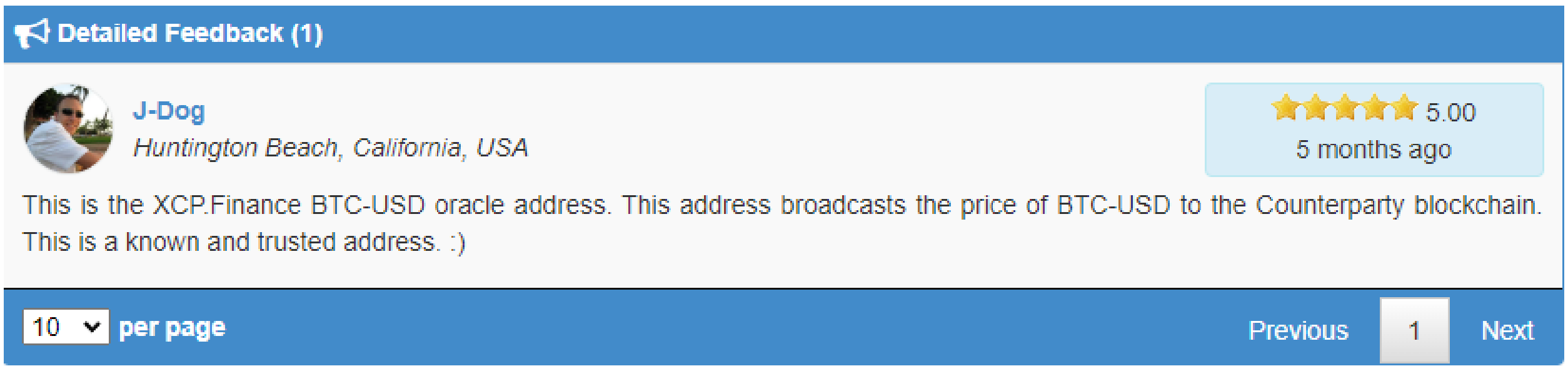

The oracle mechanism described above, developed and available from day one of Counterparty, is usable for almost any protocol application. Users could, for example, also wager on the outcome of a feed placing their bets into an escrow which is settled when the feed that they rely on passes the chosen deadline [59].

### 3.6. A provably honest oracle: Oraclize

Thomas Bertani's main interest was the use of Graphical Processing Units (GPUs) for scientific calculations. Being similar to the concept of mining, he becomes soon interested in Bitcoin when he heard about it in 2012. He had direct experience in the development and production of ASIC miners with Cointerra in 2013 and with exchanges as the founder of BitBoat, the first Italian company that allowed the purchase of bitcoin by cash.

As an innovator, he was touring to present blockchain as well as other technology updates. While studying to prepare his speech for Codemotion, an event in Milan, he came across the BitcoinWiki page started by Mike Hearn that discussed oracles as co-signers and was "fascinated by those complex conditional transactions." Then he began to advertise also those concepts at conferences.

Bertani noticed that while presenting blockchain features, the audience was highly interested in automated transactions (smart contracts) based on real-world events. Since oracles were crucial to achieving those automated transactions, he thought they would have been something huge in the short term. On this early thought, however, he realized that: "I was wrong. Oracles are still a very long-term problem; I don't think there is a convincing solution at the moment. It's a partially solved problem for very simple use cases such as price feeds".

By that time, Bertani's main concern and purpose were to find a practical solution to the problem of feeding automated transactions with real-world data. The first version of what later became Oraclize, in fact, was based on the concept of creating a pre-authorized Bitcoin transaction by partially signing it and having the oracle put the second signature when a certain condition was met. It was, however, hard to find developers to work on the project because although they understood the potential of Oraclize, they were skeptical because of script length, high costs, and network congestion. Against skeptical opinions, Bertani managed to continue the development of Oraclize, also thanks to some hackathons awards, the first of which was actually won by proposing a half-life insurance model based on Oraclize.

When Ethereum was launched in May 2015, the Oraclize interface was adapted to run on the new blockchain, with the first tests done in August directly on the main net. It was soon actively used in smart contracts as Bertani declared, "the oracle to get data from the APIs was something that got much traction. We had a peak of tens of thousands of transactions every month, to get data about all different things" [60]. The team continued the development of insurance as well as other ideas, but they soon realized that every application needed a specific solution to the oracle problem. Therefore, they decided to drop all the side projects and focus exclusively on the oracle module.

By then, both Bitcoin and Ethereum versions were live and available on the Oraclize website. For Bitcoin, there was an API with a point-and-click interface and a dedicated library. For Ethereum, instead, there was a solidity integration. Due to the already discussed problems of bitcoin conditional transactions (length, costs, congestion), bitcoin integration was eventually dismissed. It must also be noted that it never went into actual production besides being used for testing purposes. The team also considered integration with Reality Keys and Amazon Mechanical Turk as data sources, but given the scarcity of requests also, these features didn't go into production.

To expand the use of the oracle, Oraclize implemented an authenticity proof to validate the fact that the oracle behaved honestly. It was called "honesty proof" at the beginning. Still, the name was soon rebranded given the fact that being data source reliability out of oracle's control, it could have created false expectations. This new feature guarantees that the data provided matches the one drawn at the source.

Given the new implementation, the project was itself rebranded in Provable as its utility was also seen outside the blockchain domain. The idea was to exploit this feature to produce proof that could support a claim in Trusted Execution Environments (TEE). Provable was, in fact, used to create key-based attestation proof for Ledger devices.

However, the first and foremost applications that succeeded in the test phase and went officially into production were developed on Ethereum and concerned random number generation for online casinos. The advantage in Provable was that, unlike an auditing attestation service that verifies online casino platforms every now and then, the Provable engine guaranteed that every number generation was executed safely. In the blockchain field, gambling was already a trend for applications such as Satoshi dice. It already had a large user base, besides insurance and decentralized finance, that came much later.

Despite the increase in efficiency of the whole project, sadly, Provable could not keep up in popularity and traction with other modern oracles due to different business choices. Mainly the absence of a token favored those that used it for marketing purposes, eventually influencing the oracle market shares and distribution. That said, the oracle is still under development, and since its inception, it has processed millions of transactions on the Ethereum blockchain making it one of the longest-running and most widely used oracles to date.

### 3.6.1. How Oraclize worked on Bitcoin.

Oraclize in Bitcoin could be leveraged using conditional transactions and P2SH. The Bitcoin script shall include the condition (or set of conditions), the required signature to be redeemed, the data source, the outcome (or set of outcomes), and possibly an expiration date so that in case of oracle malfunction or inability of parties to signing the transactions, the funds are returned to their owners.

To better clarify the use of Oraclize on Bitcoin, an example can be taken from the protocol library that involves a bet between two agents, Alice and Bob [61].

The two agents establish that if the temperature in Milan (Italy) is above 10 degrees or if it rains from the contract is established until the next 24 hours, Bob can unlock the funds; otherwise, Alice can. They establish that the conditions are checked hourly via Wolfram Alpha until a condition is matched or the time limit elapses.

In this contract, we then have the following:

- A number of **agents**, Alice, bob, and Oraclize. A fourth agent (e.g., Carol) can be the arbitrator in case one of the others is unreachable or inactive. Otherwise, a nLockTime script can establish the refund of the money after a certain amount of time.

- Pre-established **conditions** and **outcomes**. The conditions are the temperature in Milan above ten degrees and the event of rain. In both cases, the outcome is that bob can unlock the funds. If both conditions are not verified within the timeframe, the outcome is that Alice takes the funds. Action outcomes can overlap, but conditions should not. This is in order to avoid ambiguity in the contracts and certain types of attacks. For example, if a condition is ambiguous, an oracle can select the most convenient result for selfish purposes.
- A data source, which is **Wolfram Alpha** in this case, but the parties can agree upon any other source.
- A **Timeframe** in which the contract is active.

The contract resolves when two of the three key owners sign the transaction.

Truthfully, the contract can be written and also established without the help of Oraclize, as it can directly point to a web API. The data provider may offer its signature to the data itself and also to the transaction. What Oraclize does, however, is standardize the data transfer and the authenticity proof so that any web API can be a data source for the blockchain without any adaptation from their end. The authenticity proof is also available in the case the data source does not sign the data. Leveraging Qualcomm TEE technology, Oraclize can also guarantee that the data drawn from the web API has not been manipulated. Intuitively, only data sources with SSL encryption can be utilized with Oraclize since, in the absence of this level of security, the protocol would be unable to guarantee the reliability of the data due to unforeseen man-in-the-middle attacks. An additional level of security was also developed called “ProofShield,”. With this feature, it could have been ensured that the proof of authenticity was already correctly verified if a transaction was signed. Without “ProofShield”, on a chain like Bitcoin, the verification could not have been enforced but only verified and audited manually at a later time.

Although available and working, the Bitcoin integration of Oraclize was dismissed since there were not enough requests to justify the put into production.

## 4. Discussion

This section elaborates on the findings that emerged to answer this study's research questions. The first paragraph discusses the origin of the oracle idea, the intuitions at the base of the following proposals, and the debates that emerged. The second paragraph discusses the characteristics of trust models and how those protocols addressed the conundrum of the oracle problem. The third provides an overview of the limitations of building oracles on Bitcoin and further elaborates on the passage to Ethereum/alt-chains.

### 4.1. Oracles and extrinsic data on-chain debate

As per experts’ experience, the oracle concept's first appearance came from the developer Mike Hearn that had it formalized in a BitcoinWiki post. In his interview, Hearn stated that he was not inspired by the work of someone else, but he borrowed part of the idea directly from the computer science concept of the “random oracle model.” It also emerged that the name was meant to be provisional since oracles in computer science referred to something quite the opposite to what he wanted to elaborate. Arguably, if an “oracle” is a black box that feeds a centralized machine with trusted data, Hearn's proposal of a transparent box that feeds a decentralized application with trustless data should have been addressed with a different name. However, the name is stuck to date, and the heterogeneity in blockchain oracles definitions found in [5] may also be due to this taxonomical overlap. Intuitively, suppose a developer is asked to write an “oracle” for a blockchain, and the principle of the white box is not explicitly explained. In that case, he will probably write the type of oracle learned from legacy computer science. Truthfully, both oracle types have the same finality but should work in a different way and under different logic.

Interestingly, it emerged that the word “smart contract” had also been improperly used as Nakamoto named applications built on Bitcoin just “contracts.” In its contract example, of a transaction that is executed as soon as enough signatures are placed, no particular “smart” feature emerges. They appear to be digitalized representations of ordinary contracts based on a blockchain. In this form of contract, all the parties, ideally, share the same power (keys). Considering their characteristics, in the official wiki written by Mike Hearn, they are, in fact, called “distributed contracts,” which seem to be more adherent to the original idea [22]. According to the reminiscences of Mike Hearn, the prefix “smart” started to be used a bit later. On the one hand, because it slightly resembled the concept of smart money/smart contract developed by Nick Szabo [62], and on the other hand, it was seen as necessary to alienate the “legal aura” from the word contracts. However, same as oracles, smart contracts in origin referred to a slightly different concept [62], [63].

Concerning Nakamoto idea on oracles, we cannot fully speculate on his opinion given the limited amount of available messages and posts certainly traceable to him. From some emails, however, it emerged that he was reluctant to add on Bitcoin mainchain, data different from time. His eBay-style marketplace was, in fact, proposed on a sidechain and not on the Bitcoin mainnet. However, a marketplace such as the one he proposed, requires extrinsic data on products and feedback. Still, no explanation is given on how this extrinsic data should have been fetched. It is arguable, but not provable that Nakamoto was not planning any specific data transferring system for his platform, different from the traditional ones.

Nakamoto’s vision undoubtedly influenced early developers and enthusiasts. For bitcoin core developers, in fact, extrinsic data injection in the blockchain was often seen as an improper use of the ledger [48]. Following the idea of Nakamoto, real-world applications should have been developed only through sidechains. This general mindset affected oracles’ history in many ways. Reality Keys was, in fact, developed entirely off-chain to avoid messing with Bitcoin. Truthcoin was developed as a Bitcoin sidechain to adhere to Nakamoto ideas strictly, but due to sidechain’s slow development, it is still not an active project to date. Orisi transactions were discarded, although being legitimate, eventually leading to the project's demise. Oraclize struggled to find developers due to the skepticism of large Bitcoin scripts. Finally, Counterparty was dragged into the OP_Return war, which is also said to have impacted the Ethereum launch and development.

The OP_Return war is a matter that requires further elaboration. It was always “technically possible” to add data unrelated to bitcoin transactions on the Bitcoin blockchain. Although achievable with other features, such as the one Nakamoto utilized to add the famous string ”The Times 03/Jan/2009 Chancellor on brink of second bailout for banks” [64] on the genesis block, the OP_Return, was the easiest way to perform this operation [65], [66]. Still, adding extrinsic data with OP_Return, resulted in a transaction considered unusual or non-standard. As explained in the Orisi case, non-standard transactions are transactions that, despite being perfectly valid and minable, are not relayed by ordinary Bitcoin nodes and, therefore, are unlikely to be included into blocks. However, with a compliant miner (or by mining the transaction autonomously), it was possible to add any sort of data, such as hashes, pieces of articles, song lyrics, pieces of poetry, or pieces of whitepapers [48]. There are, in fact, online repositories such as bitcoinstrings.com that keep track of all this extrinsic data on Bitcoin. Fearing network bloat and to discourage widespread adoption of this practice, the Bitcoin core developers, with 2014 version v0.9.0., reduced the OP_RETURN payload size to 40 bytes (after a pre-release testing phase at 80 bytes), which made it practical for storing a hash plus some small metadata [67], [68]. Significant on-chain data was thought to impact transaction fees and network performances negatively. However, with this update, transactions with OP_Return of 40 bytes (or less) were considered standard and relayed by nodes with default settings.

This piece of Bitcoin history is exciting since, from this study, a discrepancy emerges in how the events are described and recalled by experts in the industry. According to the official BitcoinWiki, and reliable work of literature on the Bitcoin protocol, the OP_Return operator was inserted with version v0.9.0. and directly at 40 bytes [69], [70]. OP_Return at 80 bytes was described as an early hypothesis that was soon discarded and then accepted as an improvement in February 2015 with the v0.10.0 release. This view of history, however, clashes with some information found online and what some experts recall. OP_Return operator, in fact, appears to be already part of the Bitcoin code developed by Nakamoto in 2009 [71]. As also discussed in official forums, it was leveraged as a "non-standard" feature long before 2014 [68].

Nonetheless, the main "trigger" of the OP_Return war appears to be an early release of the v0.9.0, which de facto included a standardized OP_Return operator with a payload size of 80 bytes in middle 2013. According to what was declared by Bitcoin Core developers, the 80 bytes was a random value picked for testing purposes, and 40 bytes was then sought to be a fair amount to be finally included in the official release [68]. However, since the testing release was not widely announced and advertised to avoid overuse of the experimental features, other protocols developers building on Bitcoin in 2013 were unaware that the features they were using were meant to be "provisional." Therefore, when v0.9.0. was officially released with OP_Return payload at 40 bytes, they interpreted the "slash" as a deliberate censorship attempt, resulting in a fierce debate (therefore labeled as "war") within the community [48], [49].

Furthermore, from a technical point of view, the standardization of the OP_Return promoted with 2014 v0.9.0 was de facto inserted on Bitcoin in 2013 (for testing purposes) with pull request #2738 [72]. Therefore, for developers already leveraging on OP_Return in 2013 as a standard feature, the 2014 standardization announcement was just a "lie," further fueling the harshness of the debate [73]. From a strictly technical perspective, in fact, the main change in the OP_Return with the v0.9.0 2014 official release was just the halving of the payload size [74].

According to other views, however, the OP_Return debate was an exaggeration since they saw it just as an excuse for some people to promote their alt-chains, blaming Bitcoin core developers for creating division in the community [48], [54].

Further details on the OP_Return debate are out of the scope of this research. What clearly emerges is that the discussion of whether it is right or not to inject extrinsic data into Bitcoin is a conundrum of difficult solution. If on the one hand, Nakamoto's vision is evident on the fact that Bitcoin should remain pure (of data except for time); on the other hand, inventions in history are not always used as the inventor intended. In the case of Bitcoin, however, being decentralized and maintained by the community, any network "misuse" is paid by all the nodes regardless of their approval. As some objected, however, the exponential growth of the Bitcoin blockchain is also due to an increase in its use rather than just arbitrary data injection [73], [75].

Although still unsolved, nowadays, the debate is of less interest since real-world applications are mainly built on alt-chains such as Ethereum. In the author's opinion, however, the idea of Nakamoto to keep the main chain pure and experiment on additional layers, alt-chain or sidechains, could have been a reasonable, fair take at the end.

### 4.2. Approaches to the oracle problem and trust models

Given the nascency of decentralized machines and Bitcoin constraints, approaching the oracle problem was undoubtedly a hard task for early developers. In the contracts proposals by Nakamoto, the concept of trust was completely absent, as he probably aimed for a purely trustless environment. If a contract was based on an external party, he had to sign his part before it was broadcasted, therefore excluding any possible moral hazard [30]. The case described in [21] instead shows the outcome of the contract subject to the approval of the majority of its signers, arguably a voting-based system. There is, however substantial difference between the system described by Nakamoto and the voting-based system proposed in later oracle trust models. In the voting system proposed by Nakamoto, the outcome of the contract was meant to produce effect only for its signers. Later oracle proposals instead outline oracle systems in which voters' actions produce effect both for themselves and for platform users [76], [77].

The trust model proposed by Paul Sztorc is the one that most reflects Nakamoto's idea, although extending the role of voters. His approach addresses the oracle problem, trying to eliminate the concept of a single actor as a central point of trust. Of course, the corporation proposed by Sztorc is itself a central authority, but the economic incentive is meant to prevent a possible takeover by a single actor. Since based on sidechains, the system's security is still to be evaluated. As a technology that is not yet available, it is hard to predict how it may and will affect the overall safety of the oracle.

The trust model proposed in other oracle mechanisms in this study leverages different factors to guarantee the reliability of data on-chain. Reality Keys, for example, aimed at implementing a data source that could have been widely recognized as reliable (e.g., Freebase). The reliability of the oracle protocol would have then benefitted from the trustworthiness of the data source itself. In case of mistakes due to automation, Reality Keys also included the possibility of a manual check upon the payment of a fee. Of course, this reduces the externalities of a machine failure but also introduces the chance of human failure. Trust in the very actors running the protocol is then required. The smart contract's security is trustless to a certain extent. As the contract setup is to be made directly by the clients, they are responsible for their security.

Orisi's trust model needs to be seen and analyzed under its own logic. Orisi proposes a system based on multiple data feeds of high reputation from the traditional financial world. Arguably Bitcoin was supposed to propose an alternative to the existing financial world; however, it could not offer the stability (in terms of value) guaranteed by traditional finance. Therefore, the idea of Orisi to launch a fiat-pegged stablecoin based on reliable data feeds from traditional finance is entirely coherent with its aim. A trustless data feed would not have been a logical conclusion. On the other hand, its design based on multiple oracles offers a certain degree of decentralization, which should guarantee a stable and reliable feed in case of malfunction or unavailability of some of the data sources.

The trust model offered in Oraclize found its premises in the power of technology. Leveraging on Trusted Execution Environments, it aims at guaranteeing that the data fetched at a reliable data source, such as Wolfram Alpha, has not been manipulated. Its design is undoubtedly centralized, but its features are meant to prove, in a fully auditable way, that it does not suffer from the usual weakness of a centralized source while enjoying the relative advantages. It went officially into production first for gaming applications, for which a centralized but secure design was plainly appropriate.

The oracle module available in Counterparty has quite a simple structure that is explainable by the complexity of the whole protocol. Counterparty Oracle evaluation is open to the public judgment of xchain.io users that, with their feedback, can increase the rating of the data feed. However, the power and reliability of this system are subject to the size of the active community. If the community is restricted and non-active (e.g., made of speculators), it's unlikely that those oracles are evaluated, or competition between them arises. Arguably, however, with a low TVL, the chance of manipulation is also low. In case the active community is considerably large in size, then more comments and feedback to the oracles are expected, therefore increasing their meaningfulness. However, with a higher TVL, the chance of manipulation will also increase, but arguably more oracle alternatives should also be available.

The outcome of this study suggests that the third research question cannot be answered by relying on a chronological order. Trust model design did not evolve with time but adapted almost instantaneously according to specific applications and needs. After the idea of oracles was launched by Mike Hearn in 2011 and further advertised in his 2012 London talk, a certain amount of time was required for developers to elaborate further on these new concepts and come up with a solution. As discussed by all the experts, 2013 was the year they elaborated their project to have those published with a proposal and/or a whitepaper in 2014. Therefore, all the different trust models came out almost simultaneously, de facto, weakening the hypothesis of evolution. The newborn trust models were untied from each other's and reflected their practical use and purposes.

The models analyzed in this study have their own peculiarities and uniqueness; therefore, none of these can be considered an improvement of another one. An analysis of second-generation oracles that may be made of those natives of the Ethereum blockchain could show some inspiration or improvement to those analyzed in this study. However, an investigation into these is beyond the scope of this study.

### 4.3. Limitations and difficulties of building on Bitcoin and the transition to Ethereum

It is widely known that building on Bitcoin in the early days was a difficult task; therefore also, oracles development was problematic. According to experts' opinion, the main difficulties concerned:

- The absence of developing tools and wallets
- The large size and costs of Bitcoin scripts
- Concerns about net congestion.
- Skepticism in storing extrinsic data into Bitcoin.

The existence of applications, such as Satoshi Dice or Lighthouse, supports the view that it was actually possible to build on Bitcoin, but the development was not standardized, and every developer had to find the proper workaround for their application. Although Hearn developed Bitcoin-J as a wallet aimed at being like Metamaks for Ethereum, it lacked the contribution of other developers to further build on top of it. The experience of Edmund Edgar with Runkeeper API also confirms that although it was possible to build with Bitcoin, the absence of a proper wallet and developing tools constituted a critical limitation. With developing tools and Metamask, EVM undoubtedly constituted an incentive for developers to migrate to Ethereum.

The concerns on net congestion and transaction costs were another element that further contributed to pushing real-world application development outside the Bitcoin domain. Despite the interest and prizes that Oraclize managed to obtain, he could never put the Bitcoin version into production due to low usage and the struggle to find developers. Orisi project was abandoned due to the skepticism around scripts and the inability to have their transaction mined. Their experience made clear that no application could entirely rely on non-standard scripts for its survival. Truthcoin was built following Nakamoto's advice, but to date, a working version is still unavailable due to the struggle to build a proper sidechain. Although nonfacing the issues of building directly on Bitcoin, it is suffering from the issue of having not an existing chain to be developed upon. In fact, other projects inspired by Truthcoin, such as Augur, could have been successfully launched on Ethereum already in 2016 [78]. However, complying with different standards, it is debatable if that choice constitutes an improvement of the original design.

However, the actual limit of building on Bitcoin emerges through the experience of Counterparty. Regardless of whether they were using or misusing OP_Return feature, their history shed light on the fact that building on Bitcoin was simply not "welcome" [48]. The Ethereum gas system compromise allows anyone to program any type of application as long as the proper gas fees are paid. It was born with this system, and nodes know their role and purpose. On Bitcoin, different philosophies and visions coexist, and not all the nodes/miners share the same idea. Although Bitcoin has a system of fees that varies according to the net congestion and transaction type, it was not meant to include also applications in the first place. Therefore, the payment of a fee is not necessarily a good compromise for those who wish for a light and mono-purpose chain. When Ethereum was launched then, the environment was divided into two main platforms, one of which was tormented by disputes on extrinsic data usage and block size, and another one that was cheap and full of enthusiasts building and experimenting [79], [80]. Above all, many fundings were also coming to the Ethereum platform, so it was understandable to expect a consistent migration of developers [81].

Nowadays, many improvements have been made to the Bitcoin network with the development of second layers, such as the Lightning network [82]. Ideally, those are capable of bringing the entire ecosystem built on Ethereum to the Bitcoin network. Also, advancements have been made by Blockstream concerning sidechains. It is arguable then to expect a working version in the near future. Nonetheless, due to the shift from an electronic cash system to a safe-haven asset, as a consequence of the block size war, many of those who own a significant amount of bitcoins share the philosophy of HODL (hold for dear life). Therefore, even if working decentralized applications will eventually be built on Bitcoin, skepticism emerges on the existence of a solid user base willing to spend their assets on them.

## 5. Conclusions

This study provides an overview of the Bitcoin oracles' history, from the first theoretical idea to the early practical applications till the advent of Ethereum. In the absence of dedicated literature, experts who worked on oracles in the early days were interviewed, and the information provided was enriched with the available written material found online. From the research, it emerges that the idea of an oracle mechanism came from Mike Hearn, that had it formalized in an early BitcoinWiki page. The concept was then further elaborated theoretically by other experts and then translated into actual software by a few enthusiasts. The year in which those projects were developed was 2014. All approaches to solving the oracle problem bear their peculiarities that are primarily due to the specific applications for which they were designed. The idea of a chronological evolution of trust models is instead not verified as those investigated were apparently untied from each other. Another aspect that emerges from this research is the difficulty in building oracles and, in general, applications on Bitcoin. Interestingly, the hardest to overcome were not technical difficulties. A part of the Bitcoin community was, in fact, reluctant to introduce extrinsic data on the chain due to concerns about network growth/congestion and transaction fees. The same goes for some non-standard Bitcoin scripts. The passage to Ethereum was, therefore, inevitable.

The present research contributes to academic literature filling the gap that exists from the origin of oracles on Bitcoin to modern oracles on Ethereum/alt-chains. The original concept of oracles as well as smart contracts, are clarified. The theoretical background of future academic papers can therefore build on the findings of this research. Practitioners can also benefit from this research by understanding how oracles were theorized at early stages and how they were initially adapted to different applications.

The present study also has limitations since, although data were double checked, and verified by the author, the history is described through the eyes of the experts interviewed; therefore, it can be biased by their personal views and background. Furthermore, the impossibility of interviewing Nakamoto, and given the scarcity of the retrieved material concerning his opinion/idea on oracles, the accuracy of the interpretation provided cannot be guaranteed.

Further studies can build on this one by comparing the oracles and trust models analyzed in this paper with those developed afterward on Ethereum and other alt-chains.

Appendix 1. BitcoinWiki “Contract” page, contributors, and contribution types.

| **Name and/or Pseudonym** | **First contribution** | **Last Contribution** | **Contribution type** |
|---|---|---|---|
| Mike Hearn (Mike) | The 22nd of May 2011 | The 25th of May 2014 | Creator and main contributor to the page, added the main ideas. |
| Stephen Gornick (Sgornick) | The 02nd of June 2011 | The 18th of June 2012 | Contributed to example 7: Micropayments. |
| Santacruz | 15th September 2011 | / | Minor Edit |
| II | The 07th of February 2012 | / | General punctuations to improve readability. |
| Alex Axelrod (Aaxelrod) | 8th March 2012 | / | Added Contract examples |
| Amiller | The 01st of March 2013 | The 10th of October 2013 | Included TierNolan's Protocol |
| Tumak | The 13th of June 2013 | / | Example 5: Trading across chains. |
| XertroV | The 25th of September 2013 | / | Added missing data in Example 5. |
| Peter Todd (Petertodd) | The 14th of January 2014 | / | Warned about Tx replacements |
| Ragnar | 8th April 2014 | / | Minor Edit |
| Nanotube | The 09th of April 2014 | / | Minor Edit |
| Thomas Kolinko (Kolinko) | The 09th of June 2014 | / | Added to Multiple Independent Oracles |
| Knaperek | 31th August 2014 | / | Refined terminology Example 7 |
| Topynate | The 10th of January 2015 | / | Fixed link to commit stubbing out of Tx Replacement |
| Sunnankar | 13th June 2015 | / | Added to multiple independent oracles. |
| Phantomcircuit | The 20th of June 2015 | / | Added thoughts on insecure transactions. |
| Ysangkok | 24th June 2017 | / | Added references to the lighting network |
| Belcher | 10th August 2017 | / | Minor Edit |
| Jhfrontz | The 29th of November 2018 | / | Minor Edit |
| Jonathan Cross | The 25th of February 2019 | / | Minor Edit |

**References**

[1] G. Andresen, "Bit-thereum | GavinTech," *GavinTech*, Jun. 09, 2014. http://gavintech.blogspot.com/2014/06/bit-thereum.html (accessed Jan. 21, 2023).

[2] G. Caldarelli, "Understanding the Blockchain Oracle Problem : A Call for Action," *Information*, vol. 11, no. 11, 2020, doi: 10.3390/info11110509.

[3] G. Caldarelli, *Blockchain Oracles and the Oracle Problem: A practical handbook to discover the world of blockchain, smart contracts, and oracles —exploring the limits of trust decentralization.*, 1st ed. Naples, Italy: Amazon Publishing, 2021.

[4] R. Belchior, A. Vasconcelos, S. Guerreiro, and M. Correia, "A Survey on Blockchain Interoperability: Past, Present, and Future Trends," *ACM Computing Surveys*, vol. 54, no. 8. ACM PUB27 New York, NY, pp. 1–41, Nov. 30, 2022. doi: 10.1145/3471140.

[5] G. Caldarelli, "Overview of Blockchain Oracle Research," *Futur. Internet*, vol. 14, no. 6, p. 175, Jun. 2022, doi: 10.3390/fi14060175.

[6] S. K. Ezzat, Y. N. M. Saleh, and A. A. Abdel-Hamid, "Blockchain Oracles: State-of-the-Art and

Research Directions," *IEEE Access*, vol. 10. Institute of Electrical and Electronics Engineers Inc., pp. 67551–67572, 2022. doi: 10.1109/ACCESS.2022.3184726.

[7] S. Ibba, A. Pinna, G. Baralla, and M. Marchesi, "ICOs overview: Should investors choose an ICO developed with the lean startup methodology?," *Lecture Notes in Business Information Processing*, vol. 314. pp. 293–308, 2018. doi: 10.1007/978-3-319-91602-6_21.

[8] S. Lahajnar and A. Rožanec, "Initial coin offering (ICO) evaluation model," *Invest. Manag. Financ. Innov.*, vol. 15, no. 4, pp. 169–182, 2018, doi: 10.21511/imfi.15(4).2018.14.

[9] B. Barraza, "The worth of words: How technical white papers influence ICO blockchain funding," *MIS Q. Exec.*, vol. 18, no. 4, pp. 281–285, 2019, doi: 10.17705/2msqe.00021.

[10] H. Treiblmaier, "The impact of the blockchain on the supply chain: a theory-based research framework and a call for action," *Supply Chain Manag. An Int. J.*, vol. 23, no. 6, pp. 545–559, Sep. 2018, doi: 10.1108/SCM-01-2018-0029.

[11] A. Kumar, R. Liu, and Z. Shan, "Is Blockchain a Silver Bullet for Supply Chain Management? Technical Challenges and Research Opportunities," *Decis. Sci.*, vol. 51, no. 1, pp. 8–37, Feb. 2020, doi: 10.1111/deci.12396.

[12] A. Egberts, "The Oracle Problem - An Analysis of how Blockchain Oracles Undermine the Advantages of Decentralized Ledger Systems," *SSRN Electron. J.*, 2017, doi: 10.2139/ssrn.3382343.

[13] J. Frankenreiter, "The Limits of Smart Contracts," *J. Institutional Theor. Econ. JITE*, vol. 175, no. 1, pp. 149–162, 2019, doi: 10.1628/jite-2019-0021.

[14] M. Damjan, "The interface between blockchain and the real world," *Ragion Prat.*, vol. 2018, no. 2, pp. 379–406, 2018, doi: 10.1415/91545.

[15] A. Pasdar, Z. Dong, and Y. C. Lee, "Blockchain Oracle Design Patterns," pp. 1–25, Jun. 2021, [Online]. Available: http://arxiv.org/abs/2106.09349

[16] H. Al-Breiki, M. H. U. Rehman, K. Salah, and D. Svetinovic, "Trustworthy Blockchain Oracles: Review, Comparison, and Open Research Challenges," *IEEE Access*, vol. 8, pp. 85675–85685, 2020, doi: 10.1109/ACCESS.2020.2992698.

[17] S. Eskandari, M. Salehi, W. C. Gu, and J. Clark, "SoK: Oracles from the Ground Truth to Market Manipulation," *Proc. Under Rev. (May 2021)*, vol. 1, no. 1, Jun. 2021, [Online]. Available: http://arxiv.org/abs/2106.00667

[18] B. Liu, P. Szalachowski, and J. Zhou, "A First Look into DeFi Oracles," *arXiv*, no. March, May 2020, [Online]. Available: http://arxiv.org/abs/2005.04377

[19] A. Pasdar, Y. C. Lee, and Z. Dong, "Connect API with Blockchain: A Survey on Blockchain Oracle Implementation," *ACM Comput. Surv.*, Aug. 2021, doi: 10.1145/3567582.

[20] A. Beniiche, "A study of blockchain oracles," *arXiv*, pp. 1–9, 2020.

[21] S. Nakamoto, "Re: Open sourced my Java SPV impl | Satoshi's Archive," *Bitcoin.com*, Mar. 09, 2011. https://www.bitcoin.com/satoshi-archive/emails/mike-hearn/13/ (accessed Nov. 25, 2022).

[22] M. Hearn, "Contracts," *BitcoinWiki*, 2011. https://en.bitcoin.it/w/index.php?title=Contract&oldid=13637 (accessed Dec. 02, 2022).

[23] Furunodo, "Contracts," *BitcoinWiki*, 2020. https://en.bitcoin.it/w/index.php?title=Contract&oldid=67871 (accessed Dec. 12, 2022).

[24] J. Southurst, "Apple Removes Blockchain Bitcoin Wallet Apps From its App Stores," *CoinDesk*, Feb. 06, 2014. https://www.coindesk.com/markets/2014/02/06/apple-removes-blockchain-bitcoin-

wallet-apps-from-its-app-stores/ (accessed Nov. 21, 2022).

[25] S. Nakamoto, “Re: Lack of chargeback support | Satoshi’s Archive,” *Nakamoto Email*, Apr. 27, 2009. https://www.bitcoin.com/satoshi-archive/emails/mike-hearn/8/ (accessed Nov. 25, 2022).

[26] S. Nakamoto, “Escrow,” *Bitcointalk.org*, Aug. 07, 2010. https://bitcointalk.org/index.php?topic=750.0 (accessed Feb. 20, 2023).

[27] Sebastian, “Bitcoin secure chargebacks (with votes)?,” *Bitcointalk.org*, Mar. 2011. https://bitcointalk.org/index.php?topic=4856.0 (accessed Feb. 20, 2023).

[28] S. Nakamoto, “Re: 2 Open sourced my Java SPV impl | Satoshi’s Archive,” *Bitcoin.com*, Mar. 09, 2011. https://www.bitcoin.com/satoshi-archive/emails/mike-hearn/14/#selection-25.4239-25.4589 (accessed Nov. 25, 2022).

[29] Appamatto, “BitDNS and Generalizing Bitcoin,” *Bitcointalk.org*, Nov. 15, 2010. https://bitcointalk.org/index.php?topic=1790.0 (accessed Feb. 20, 2023).

[30] S. Nakamoto, “Re: Holding coins in an unspendable state for a rolling time window | Satoshi’s Archive,” *Bitcoin.com*, Apr. 20, 2011. https://www.bitcoin.com/satoshi-archive/emails/mike-hearn/15/ (accessed Nov. 25, 2022).

[31] A. Piscitello, “Implementing External State Contracts - Feedback Requested,” *Bitcointalk.org*, Jul. 23, 2013. https://bitcointalk.org/index.php?topic=260898.0 (accessed Jan. 03, 2023).

[32] J. Southurst, “Reality Keys: Bitcoin’s Third-Party Guarantor for Contracts and Deals,” *Coindesk.com*, 2014. https://www.coindesk.com/markets/2014/01/17/reality-keys-bitcoins-third-party-guarantor-for-contracts-and-deals/ (accessed Jan. 06, 2023).

[33] E. Edgar and C. Delrey, “Realitykeysdemo.py,” *Github.com*, 2014. https://github.com/edmundedgar/realitykeys-examples/blob/master/realitykeysdemo.py (accessed Jan. 06, 2023).

[34] S. Schaefer, “Intrade Closes To U.S. Bettors, Bowing To Pressure From Regulators,” *Forbes.com*, Nov. 27, 2012. https://www.forbes.com/sites/steveschaefer/2012/11/27/cftc-takes-aim-at-intrade-files-suit-going-after-prediction-market/?sh=7590e5ae66b9 (accessed Dec. 01, 2022).

[35] C. Isidore, “Intrade shut down due to financial probe,” *CNN Business*, Mar. 11, 2013. https://money.cnn.com/2013/03/11/investing/intrade-shutdown/index.html (accessed Dec. 01, 2022).

[36] A. Back *et al.*, “Enabling Blockchain Innovations with Pegged Sidechains,” 2014.

[37] R. Hanson, “Logarithmic Markets Coring Rules for Modular Combinatorial Information Aggregation,” *J. Predict. Mark.*, vol. 1, no. 1, pp. 3–15, 2012, doi: 10.5750/jpm.v1i1.417.

[38] G. Angeris and T. Chitra, “Improved Price Oracles: Constant Function Market Makers,” *SSRN Electron. J.*, pp. 80–91, 2020, doi: 10.2139/ssrn.3636514.

[39] P. Sztorc, “Truthcoin Peer-to-Peer Oracle System and Prediction Marketplace,” *bitcoinhivemind.com*, Dec. 14, 2015. https://bitcoinhivemind.com/papers/truthcoin-whitepaper.pdf (accessed Feb. 15, 2023).

[40] “salience noun - Definition, pictures, pronunciation and usage notes | Oxford Advanced Learner’s Dictionary at OxfordLearnersDictionaries.com,” *oxfordlearnersdictionaries.com*. https://www.oxfordlearnersdictionaries.com/definition/english/salience (accessed Feb. 15, 2023).

[41] S. Bistarelli, I. Mercanti, and F. Santini, “An Analysis of Non-standard Transactions,” *Front. Blockchain*, vol. 2, p. 7, Aug. 2019, doi: 10.3389/FBLOC.2019.00007.

[42] T. C. Schelling, “The strategy of conflict . Public Domain , Google-digitized,” p. 309, 1960, Accessed: Feb. 13, 2023. [Online]. Available: https://www.hup.harvard.edu/catalog.php?isbn=9780674840317

[43] J. Warren, “Bitmessage: A Peer-to-Peer Message Authentication and Delivery System,” 2012, Accessed: Nov. 24, 2022. [Online]. Available: www.Bitmessage.org

[44] T. Kolinko, G. Pstrucha, and K. Kucharski, “Orisi Whitepaper,” *GitHub*, 2014. https://github.com/orisi/wiki/wiki/Orisi-White-Paper (accessed Sep. 06, 2022).

[45] J. Willett, “The Second Bitcoin Whitepaper,” *White Pap.*, vol. 5, 2013, [Online]. Available: https://github.com/bitsblocks/mastercoin-whitepaper/blob/master/index.md

[46] CryptoApisTeam, “What is Omni Layer and How Does It Work?,” *Cryptoapis*, Jun. 09, 2022. https://cryptoapis.io/blog/89-what-is-omni-layer-and-how-does-it-work (accessed Jan. 13, 2023).

[47] L. Shin, “Here’s The Man Who Created ICOs And This Is The New Token He’s Backing,” *Forbes.com*, Sep. 21, 2017. https://www.forbes.com/sites/laurashin/2017/09/21/heres-the-man-who-created-icos-and-this-is-the-new-token-hes-backing/?sh=3d1d65411839 (accessed Jan. 13, 2023).

[48] B. Research, “The OP_Return Wars of 2014 – Dapps Vs Bitcoin Transactions,” *blog.bitmex.com*, 2022. https://blog.bitmex.com/dapps-or-only-bitcoin-transactions-the-2014-debate/ (accessed Jan. 12, 2023).

[49] bchworldorder, “‘A few months after the Counterparty developers started using OP_RETURN, bitcoin developers decreased the size of OP_RETURN from 80 bytes to 40 bytes. The sudden decrease in the size of the OP_RETURN function stopped networks launched on top of bitcoin from operating properly.’ : btc,” *Reddit.com*, 2018. https://www.reddit.com/r/btc/comments/80ycim/a_few_months_after_the_counterparty_developers/ (accessed Jan. 19, 2023).

[50] D. Bradbury, “Developers Battle Over Bitcoin Block Chain,” *Coindesk.com*, 2014. https://www.coindesk.com/markets/2014/03/25/developers-battle-over-bitcoin-block-chain/ (accessed Jan. 19, 2023).

[51] L. Dashjr, “[ANN][XCP] Counterparty - Pioneering Peer-to-Peer Finance - Official Thread,” *Bitcointalk.org*, Mar. 21, 2014. https://bitcointalk.org/index.php?topic=395761.msg5817170#msg5817170 (accessed Jan. 15, 2023).

[52] Historian1111, “Blockstream Co-founder Luke-jr banning Mastercoin and Counterparty transactions, adding blacklists to Gentoo bitcoind by default.,” *reddit.com*, 2015. https://www.reddit.com/r/Bitcoin/comments/2pfxak/blockstream_cofounder_lukejr_banning_mastercoin/ (accessed Jan. 17, 2023).

[53] Insette, “Your best pitch for Decred,” *old.reddit.com*, 2018. https://old.reddit.com/r/decred/comments/6wxueo/your_best_pitch_for_decred/dmcer4d/ (accessed Jan. 19, 2023).

[54] M. I. Moneyist, “The OP_Return war ‘debunked,’” *Twitter.com*, 2019. https://twitter.com/notgrubles/status/1187470076833697794 (accessed Jan. 14, 2023).

[55] V. Buterin, “The very earliest versions of ETH protocol,” *Twitter.com*, Nov. 12, 2017.

[56] E. Muratov, “Bitcoin Minimalism: Counterparty to Talk with Bitcoin in Ethereish,” *ForkLog*, 2016. https://web.archive.org/web/20170623073803/http:/forklog.net/bitcoin-minimalism-counterparty-to-talk-with-bitcoin-in-ethereish/ (accessed Jan. 20, 2023).

[57] K. Pani KV, “How to create the Meta Chain,” *blog.sap.com*, 2014. https://blogs.sap.com/2014/01/06/how-to-create-the-meta-chain/ (accessed Jan. 16, 2023).

[58] D. Weller, “Decoding a transaction,” *Github.com*, 2016. https://github.com/tokenly/counterparty-spec/blob/master/spec/02-decoding.md (accessed Jan. 24, 2023).

[59] D. Weller, I. Zuber, and Chiguiretor, “Protocol Specification | Counterparty,” *Github.com*, 2019. https://github.com/CounterpartyXCP/Documentation/blob/master/Developers/protocol_specification.md (accessed Jan. 25, 2023).

[60] Etherscan, “Oraclize address transactions,” *Etherscan.io*, 2023. https://etherscan.io/address/0x26588a9301b0428d95e6fc3a5024fce8bec12d51#analytics (accessed Jan. 30, 2023).

[61] D-Nice, “GitHub - provable-things/oraclize-lib: Oraclize node.js library,” *github.com*, May 12, 2017. https://github.com/provable-things/oraclize-lib (accessed Jan. 27, 2023).

[62] N. Szabo, “Smart Contracts,” *Personal Blog*, 1994. https://www.fon.hum.uva.nl/rob/Courses/InformationInSpeech/CDROM/Literature/LOTwinterschool2006/szabo.best.vwh.net/smart.contracts.html (accessed Feb. 08, 2023).

[63] Szabo Nick, “Formalizing and securing relationships on public networks,” 1997. https://journals.uic.edu/ojs/index.php/fm/article/view/548 (accessed Feb. 15, 2020).

[64] C. Murray, “The mystery of the Genesis block - CoinGeek,” *Coingeek.com*, Dec. 07, 2021. https://coingeek.com/the-mystery-of-the-genesis-block/ (accessed Feb. 07, 2023).

[65] N. O’Dell, “op return - What was the very initial value of OP_RETURN? - Bitcoin Stack Exchange,” *bitcoin.stackexchange.com*, Dec. 29, 2016. https://bitcoin.stackexchange.com/questions/50414/what-was-the-very-initial-value-of-op-return (accessed Feb. 06, 2023).

[66] M. Bartoletti and L. Pompianu, “An analysis of bitcoin OP_RETURN metadata,” *Lect. Notes Comput. Sci. (including Subser. Lect. Notes Artif. Intell. Lect. Notes Bioinformatics)*, vol. 10323 LNCS, pp. 218–230, 2017, doi: 10.1007/978-3-319-70278-0_14/FIGURES/3.

[67] Seandotau, “OP_RETURN 40 TO 80 BYTES,” *talkcrypto.org*, 2016. https://www.talkcrypto.org/blog/2016/12/30/op_return-40-to-80-bytes/ (accessed Jan. 23, 2023).

[68] G. Andresen, “Relay OP_RETURN data TxOut as standard transaction type.,” *Github.com*, 2013. https://github.com/bitcoin/bitcoin/pull/2738 (accessed Jan. 23, 2023).

[69] A. M. Antonopoulos, *Mastering Bitcoin: Programming the Open Blockchain*, 2nd ed. O’Reilly, 2017.

[70] “OP_RETURN - Bitcoin Wiki.” https://en.bitcoin.it/wiki/OP_RETURN (accessed Feb. 07, 2023).

[71] non-github-bitcoin, “2009 Bitcoin Code,” *github.com*, Aug. 30, 2009. https://github.com/bitcoin/bitcoin/blob/4405b78d6059e536c36974088a8ed4d9f0f29898/script.cpp#L170 (accessed Feb. 07, 2023).

[72] G. Andresen, “Core Development Update #5,” *Bitcoinfoundation.org*, Oct. 24, 2013. https://web.archive.org/web/20131024212741/https://bitcoinfoundation.org/blog/?p=290 (accessed Feb. 09, 2023).

[73] T. Swanson, “Bitcoin Hurdles: the Public Goods Costs of Securing a Decentralized Seigniorage Network which Incentivizes Alternatives and Centralization,” *SSRN Electron. J.*, no. November 2008, pp. 1–52, 2014.

[74] G. Andresen, “script: reduce OP_RETURN standard relay bytes to 40 by jgarzik · Pull Request #3737 · bitcoin/bitcoin · GitHub,” *github.com*, Feb. 26, 2014. https://github.com/bitcoin/bitcoin/pull/3737/files (accessed Feb. 10, 2023).

[75] E. Strehle and F. Steinmetz, “Dominating OP Returns: The Impact of Omni and Veriblock on Bitcoin,”

*J. Grid Comput.*, vol. 18, no. 4, pp. 575–592, Dec. 2020, doi: 10.1007/S10723-020-09537-9/METRICS.

[76] J. Adler, R. Berryhill, A. Veneris, Z. Poulos, N. Veira, and A. Kastania, “Astraea: A Decentralized Blockchain Oracle,” in *2018 IEEE International Conference on Internet of Things (iThings) and IEEE Green Computing and Communications (GreenCom) and IEEE Cyber, Physical and Social Computing (CPSCom) and IEEE Smart Data (SmartData)*, Jul. 2018, pp. 1145–1152. doi: 10.1109/Cybermatics_2018.2018.00207.

[77] H. Huilgolkar, “Razor Network : A decentralized oracle platform,” 2021. https://razor.network/whitepaper.pdf (accessed Feb. 18, 2021).

[78] J. Peterson, J. Krug, M. Zoltu, A. K. Williams, and S. Alexander, “Augur: a decentralized oracle and prediction market platform,” *arXiv*, pp. 1–16, Jan. 2015, doi: 10.13140/2.1.1431.4563.

[79] C. Russo, “Sale of the Century: The Inside Story of Ethereum’s 2014 Premine - CoinDesk,” *coindesk.com*, Jul. 11, 2020. https://www.coindesk.com/markets/2020/07/11/sale-of-the-century-the-inside-story-of-ethereums-2014-premine/ (accessed Feb. 12, 2023).

[80] V. Buterin, “Launching the Ether Sale | Ethereum Foundation Blog,” *blog.ethereum.org*, Jul. 22, 2014. https://blog.ethereum.org/2014/07/22/launching-the-ether-sale (accessed Feb. 12, 2023).

[81] Cryptopedia Staff, “Initial Coin Offerings: The Ethereum ICO Boom | Gemini,” *www.gemini.com*, Mar. 10, 2022. https://www.gemini.com/cryptopedia/initial-coin-offering-explained-ethereum-ico (accessed Feb. 12, 2023).

[82] J. Poon and T. Dryja, “The Bitcoin Lightning Network: Scalable Off-Chain Instant Payments,” 2016. Accessed: Feb. 12, 2023. [Online]. Available: https://lightning.network/lightning-network-paper.pdf